\documentclass[11pt,a4paper]{article}
\usepackage{amssymb}
\usepackage{amsmath}
\usepackage{graphicx}
\usepackage{subfig}
\usepackage[english]{babel}
\newcommand{\be}{\begin{equation}}
\newcommand{\ee}{\end{equation}}
\newcommand{\bea}{\begin{align}}
\newcommand{\eea}{\end{align}}
\newcommand{\nn}{\nonumber}

\newcommand{\D}{{\cal D}}
\newcommand{\F}{{\cal F}}
\newcommand{\GG}{{\cal G}}
\newcommand{\cL}{{\cal L}}

\renewcommand{\O}{{\cal O}}

\newcommand{\V}{{\cal V}}

\newcommand{\hD}{{\hat \D}}

\renewcommand{\d}{\delta}

\newcommand{\rmd}{{\rm d}}

\newcommand{\bt}[1]{{\bar t}}
\newcommand{\ts}{\textstyle}

\newcommand{\half}{{\ts \frac{1}{2}}}
\newcommand{\quar}{{\ts \frac{1}{4}}}
\newcommand{\pr}{\partial}
\newcommand{\vep}{\varepsilon}
\newcommand{\vphi}{\varphi}

\newcommand{\tu}{\tilde u}
\newcommand{\tv}{\tilde v}

\newcommand{\hf}{{\hat f}}
\newcommand{\hh}{{\hat h}}
\newcommand{\hv}{{\hat v}}
\newcommand{\hbe}{{\hat \beta}}
\newcommand{\hga}{{\hat \gamma}}

\csname @addtoreset\endcsname{equation}{section}
\begin{document}

\begin{titlepage}
\thispagestyle{empty}
\begin{flushright}
\small
DAMTP-2008-116\\
arXiv:0901.0450 [hep-th]\\
\date \\
\normalsize
\end{flushright}

\vskip 3cm
\centerline{\LARGE \bf{Reparameterisation Invariance and RG equations:}}
\vskip 6pt
\centerline{\LARGE \bf{Extension of the Local Potential Approximation}}

\vskip 3cm

\centerline{H. Osborn\footnote{ho@damtp.cam.ac.uk} and 
D.E. Twigg\footnote{det28@cam.ac.uk}} 
\vskip 1cm
\centerline{Department of Applied Mathematics and Theoretical Physics,} 
\centerline{Wilberforce Road, Cambridge, CB3 0WA, England}


\begin{abstract}
Equations related to the Polchinski version of the exact renormalisation
group equations for scalar fields which extend the local potential approximation 
to first order in a derivative expansion, and which maintain reparameterisation 
invariance, are postulated. Reparameterisation invariance
ensures that the equations determine the anomalous dimension $\eta$
unambiguously and the equations are such that the result is exact to
${\rm O}(\vep^2)$ in an $\vep$-expansion for any multi-critical fixed point.
It is also straightforward to determine $\eta$ numerically. When the 
dimension $d=3$ numerical results for a wide range of critical exponents 
are obtained in theories with  $O(N)$ symmetry, for various $N$ and for a 
ranges of $\eta$, are obtained within the local potential approximation. 
The associated $\eta$, which follow from the derivative approximation
described here, are found for various $N$. The large $N$ limit of the equations 
is also analysed. A corresponding discussion is also given in a perturbative RG 
framework and scaling dimensions for derivative operators are calculated to 
first order in $\vep$.

\end{abstract}

\vfill ${~~~}$ \newline

\end{titlepage}

\setcounter{footnote}0
\section{Introduction}

Exact functional renormalisation group flow equations 
\cite{Wilson,Wegner,Wilson2,Polchinski,Bagnuls,Wett,Jan,Bertrand} allow, at 
least for  scalar field theories, the possibility of a non perturbative 
analysis of fixed points and determination of critical exponents which 
control the RG flow near any fixed point. In all such equations
there is a cut off function $K(p^2)$ which is essentially arbitrary save
for $K(0)=1$ and vanishing sufficiently rapidly as $p^2 \to \infty$. Any
physical results, such as precise values for exponents, should be independent
of the cut off although it may be feasible to optimise over different cut 
off functions \cite{LitimO}. The exact RG flow equations are hard to handle 
except
in some truncation or expanding in perturbation theory. The local potential
approximation (LPA) neglects the spatial dependence of the fields $\phi$ and
reduces the effective action $S[\phi]$ from a highly non trivial functional
to a simple function $V(\phi)$ and the RG flow equations become a non linear
differential equation for $V$ of the form $\dot V_t= F(V_t,V_t{\!}',V_t{\!}'')$ 
where $\dot V$ denotes the derivative with respect to $t=-\log \Lambda$, with 
$\Lambda$ the cut off scale (the equations are invariant under rescalings of 
$\Lambda$). The potential RG flow fixed points, $V_t \to V$ as $t\to \infty$,
are determined by requiring smooth solutions for all $\phi$ of 
$F(V,V',V'')=0$. The critical exponents describing the RG flow in the 
neighbourhood of the fixed point, $V_t = V + \sum_r \epsilon_r \, 
e^{\lambda_r t} v_r$, may then be calculated by finding the eigenvalues 
$\lambda_r$, and eigenfunctions $v_r$, for a corresponding linear differential
operator depending on the fixed point solution for $V$. For all $\lambda_r >0$
the associated operators are relevant and it is necessary to tune 
$\epsilon_r =0$ to attain the fixed point under RG flow. 

When applied to the Polchinski 
RG equation \cite{Polchinski}, for which  $F(V,V',V'')$ has a very simple 
quadratic form, the LPA has the virtue that all dependence of the cut off
can be removed by rescalings of $V$ and $\phi$. Although rather crude the
LPA is compatible with the global features of RG flow since, in cases that
have been investigated, it realises the same fixed points as are present in the
full quantum field theory for scalar fields that are found by other techniques
(this is not manifestly true for more complicated theories with gauge fields
and fermions \cite{Dist}).

Despite describing the essential features of the landscape of critical
points for scalar theories the LPA has, nevertheless, many limitations. 
In particular it is not possible
to consistently determine $\eta$, the anomalous dimension for $\phi$. In 
theories with dimension $d=3$, $\eta$ is generally small but clearly results
for critical exponents must then have an error of at least ${\rm O}(\eta)$,
although results when the LPA is applied to different RG flow equations differ
in general by rather more than this.  Attempts to go beyond the LPA usually 
invoke an expansion in terms of derivatives of $\phi$ 
\cite{Golner,Morris1,Ball,Morris2,Com,Morris3,Bervillier,Canet}. 
To first order this introduces 
a function $Z(\phi)$ which is the coefficient of $(\pr \phi)^2$ in an expansion
of the effective action (for multi-component fields $\phi_i$ this becomes a 
symmetric tensor $Z_{ij}(\phi))$. $Z$ and $V$ obey coupled equations which in
principle allow $\eta$ to be determined by requiring non singular solutions
for both $Z$ and $V$. However the dependence on the cut off becomes more severe 
in the derivative expansion. Applied to the
Polchinski equation there are two constants $A,B$ which are essentially 
arbitrary \cite{Ball}. Apart from this arbitrariness the results also depend 
on the value chosen for $Z(0)$ in solving the coupled equations \cite{Com}.

Exact RG equations, without approximations, are invariant under
reparameterisations, including rescalings, of the fundamental fields 
\cite{Wilson3}. This property ensures that  the full equations have
a line of physically equivalent fixed points which may be parameterised by 
different values of $Z(0)$, \cite{Com,MorrisRev,Bagnuls}. 
Physical results, such as $\eta$, are independent of where on this line 
the fixed point solution is chosen. As a consequence of
the line of equivalent fixed points the calculated exponents must include
one which is exactly zero. The corresponding marginal operator is redundant,
essentially one which vanishes on the equations of motion. 

In the context of the Polchinski RG equation it was shown, for arbitrary 
dimensions $d$, in \cite{DO} that for any local operator $\O$, such that 
$\int \rmd^d x \, \O$ gives an eigen-operator represented by an eigenfunction 
for the linearised equations with critical exponent $\lambda_\O$,
then it is possible to construct associated redundant operators with exponents
\be
\lambda = \lambda_\O - \half ( d + 2 -\eta ) - 2m \, , \qquad 
m=0,1,2, \dots \, ,
\label{red}
\ee
irrespective of any particular choice of a smooth cut off function.
Furthermore the operator $\phi$ is a local operator determining an eigenfunction
with $\lambda_\phi=\half ( d + 2 -\eta )$ and hence, applying \eqref{red} with 
$m=0$, this directly shows that $\lambda=0$ is a possible eigenvalue whose 
eigenfunction generates the marginal operator necessary for reparameterisation 
invariance.

Although reparameterisation invariance is a property of the full non linear
RG equations it is generally lost in approximations such as the derivative 
expansion. There is no longer a fluctuation eigenfunction  with $\lambda=0$ 
exactly.
Here we heuristically construct equations for $V(\phi),Z(\phi)$, and also for
multi-component generalisations, which maintain these desirable features. 
The equation for $V$ remains the same as in the LPA except for the introduction
of $\eta$. The associated equation for $Z$ depends on the solution for $V$
and determines $\eta$. Using an appropriate scalar product an integral
expression may be found which may be used to find $\eta$ in a fashion which
is manifestly independent of $Z(0)$. The eigenvalues for the
corresponding differential operator are in accord with \eqref{red} when $m=0$
and the zero mode eigenfunction can also be found explicitly.

For the purposes of comparison we also discuss results for derivative 
operators of the form $G_{ij}(\phi)\pr^\mu\phi_i \pr_\mu \phi_j$ using 
standard perturbation theory techniques and the $\vep$-expansion to obtain 
results for the anomalous dimensions at the fixed point to first order in 
$\vep$. For such derivative operators it is necessary to take account
of mixing with scalar operators $F(\phi)$ with the same dimension
but there are also additional constraints on the associated $\beta$-functions.
Keeping only contributions just to first order in the coupled $(F,G_{ij})$ then an 
infinitesimal variation $\delta \phi_i = v_i(\phi)$, for non-linear 
$v_i(\phi)$, in the lagrangian is equivalent to corresponding changes in 
$(F,G_{ij})$. This leads to identities which show that the scaling dimensions 
satisfy relations of the same form as in \eqref{red} in general in a 
perturbative framework. We are also able to determine the 
scale dimensions to ${\rm O}(\vep)$ at each of the multi-critical
fixed points for scalar theories when $O(N)$ symmetry is imposed. Although
such operators are irrelevant as far as RG flows they are of course of interest
in determining the spectrum of operators and scale dimensions in the theory
at its critical points.

In this paper we describe in the next section results for the simplest case
of a single component field which corresponds to the Ising model and has been 
much discussed previously. For $d=3$ the equations are solved numerically
and the associated eigenvalues determined for various values of $\eta$. The
appropriate value of $\eta$ necessary for a non singular solution of the
$Z$-equation is also found. In section 3, we extend the discussion to 
multi-component fields, imposing $O(N)$ symmetry so that simple equations, of
similar form to those considered in section 2, are obtained. The
eigenfunctions are then $O(N)$ tensors. The irreducible representations
are given by  symmetric traceless tensors of rank $l$ and 
the corresponding eigenvalues depend also om $l$. Numerical results are then 
given for various $N$ and $l$. In section 4 we show how these equations may be
solved in an $\vep$-expansion recovering perturbative results at the various
possible non-trivial multi-critical fixed points as the dimension $d$ is 
reduced. In section 5 we consider perturbatively the usual $\beta$-functions 
in a loop expansion, extending results obtained in the single component case 
in \cite{DO}. In section 6 these results are extended to derivative operators 
and mixing effects
taken into account. In a conclusion we make some more general remarks
concerning the status of the equations discussed in this paper. Although
they have been motivated by requiring that they share general properties
of the exact RG equations they serve to show how these may still be
maintained in quite simple approximations. Various calculational details
are relegated to four appendices. In appendix A we show how
the equations can be solved for large $N$ and a formula for $\eta$ to
${\rm O}(N^{-1})$ obtained which is quite close to the exact large $N$ result.
In appendix B we give some details of the perturbative results for
$\beta$-functions that are used in sections 5 and 6. In appendix C we
give a general discussion using dimensional regularisation of the consequences
of invariance of the regularised theory under variations $\delta \phi_i
= - v_i(\phi)$.  In the final appendix D we give some details of the 
nearest singularities that are found numerically when the solution of the 
local potential approximation for the Polchinski equation
is extended to the complex plane.

\section{Equations for a Single Component Field}

It is simplest to consider first a single scalar field $\phi$ corresponding
to the universality class for the Ising model.
At a fixed point the equation for $V(\phi)$ is
\be
V''(\phi) -\half(d-2 + \eta )\, \phi V'(\phi) -V'(\phi)^2+dV(\phi)  =0 \, .
\label{LPA}
\ee
This is just the standard LPA for the Polchinski equation including
the anomalous dimension $\eta$. In general this is set to zero as there is
no mechanism to determine this from \eqref{LPA}.
The two trivial solutions of \eqref{LPA} are $V(\phi)=0$, for
the Gaussian fixed point, and $V(\phi)= \quar(2-\eta)( \phi^2 - \frac{2}{d})$, 
for the high temperature fixed point.
Non trivial solutions even in $\phi$, so that $V'(0)=0$, which are non
singular for all $\phi$ and  
\be
V(\phi)\sim \quar(2-\eta)\, \phi^2 + A \, \phi^{\frac{2d}{d+2-\eta}} \quad
\mbox{for large} \quad \phi \, ,
\ee 
depend on a precise choice for $V(0)=k$ which then determines $A$.
Such solutions appear whenever $d$ is reduced below $2n/(n-1)$ for $n=2,3,\dots$
\cite{Felder,Lima}. The critical exponents are then determined from the eigenvalue
equation
\be
\D f(\phi) = (\lambda - d ) f(\phi) \, , 
\label{eig}
\ee
with the differential operator
\be
\D = \frac{\rmd^2}{\rmd \phi^2} -\half(d-2 + \eta )\, \phi \frac{\rmd}{\rmd \phi}
- 2 V'(\phi) \, \frac{\rmd}{\rmd \phi} \, .
\label{diff}
\ee 
It is easy to see, using \eqref{LPA}, that
\be
\D V'(\phi) = - \half ( d+2 - \eta) V'(\phi) \, , \qquad
\D \phi = - \half ( d -2 + \eta) \phi - 2  V'(\phi) \, , 
\ee
and hence we may construct two exact odd eigenfunctions
\be
f_\phi(\phi) = \half ( 2-\eta) \phi - V'(\phi) \, , \ \ \lambda_\phi =
\half(d+2-\eta) \, , 
\qquad f_R (\phi) =  V'(\phi) \, , \ \ \lambda_R =\half(d - 2 + \eta) \, ,
\label{exact}
\ee
$f_R(\phi)$ corresponds to a redundant operator with $\lambda_R$ given
by \eqref{red} with $\O$ the identity operator, which corresponds to
the solution of \eqref{eig} $f(\phi)=1$ with $\lambda = d$.

It is important for our later discussion to recognise that $\D$ in \eqref{diff} is 
hermitian with  respect to a scalar product defined by
\be
\langle f ,g \rangle = \int \rmd \phi \, e^{- \frac{1}{4}(d-2+\eta)\phi^2 
- 2V(\phi)} \, f(\phi) g(\phi) \, ,
\label{prod}
\ee
so that
\be
\langle f , \D g \rangle = \langle\D f ,g \rangle \, .
\ee

Extending the RG equations to $Z(\phi)$ we propose that, in conjunction with
\eqref{LPA}, the associated equation at the fixed point
\begin{align}
Z''(\phi) -\half(d-2 + \eta )\, \phi Z'(\phi) - 2V'(\phi)Z'(\phi)
-2V''(\phi) Z(\phi)  \nn \\ 
 = \big ( \D - 2 V''(\phi) \big ) Z(\phi) =
\eta - \frac{2d}{d+2} \, V''(\phi)^2  \, .
\label{Zeq}
\end{align}
Together with \eqref{LPA} this satisfies reparameterisation invariance
so that  $\eta$ is independent of the particular initial $Z(0)$,
unlike the case for other analogous derivative expansion equations.
To ensure non singular solutions for all $\phi$ requires only a special choice
for $\eta$. 
Asymptotically, for large $\phi$,  solutions of \eqref{Zeq} have the form
\be
Z(\phi) \sim \frac{d(2-\eta)}{2(d+2)} - \frac{\eta}{2-\eta} 
+ C \, \phi^{-\frac{2(2-\eta)}{d+2-\eta}} \, .
\label{asymZ}
\ee
In general the value of the asymptotic constant $C$ depends on $Z(0)$.

The proposed $Z$-equation \eqref{Zeq} is similar to the
derivative expansion result in \cite{Ball}. It differs in that the 
coefficient of the $V''Z$ term is 2, rather than 4, and that on the right
hand side there is a definite coefficient $\frac{2d}{d+2}$ rather than an
essentially arbitrary cut off dependent constant (in terms of the equations in
\cite{Ball} we are taking  for the cut dependent constant $B$ the 
precise value $\frac{d}{d+2}$). 
In respect of these terms  \eqref{Zeq} is identical with
an analogous equation obtained in \cite{DO} using an expansion in terms
of scaling fields which is similar in spirit to the derivative expansion. In the
scaling field approach the corresponding 
coefficient is determined precisely essentially by those divergencies in
two point amplitudes which are universal, i.e. renormalisation scheme independent. 
If $G(x)$ is the cut off dependent propagator, then we have for the following
products
\be
\frac{\pr}{\pr t} G(x)^n \sim -n a_n \delta^d(x) \, , \quad
\frac{\pr}{\pr t} G(x)^{2n-1} \sim -(2n-1) b_n \, \pr^2\delta^d(x) \, ,
\label{Feyn}
\ee
for
\be
d=d_n = \frac{2n}{n-1} \, , 
\ee
such that  $a_n,b_n$ are constants, independent of the cut off function and
depending only on the large $x$ behaviour of $G(x)$. According to the results
obtained in \cite{DO}
\be 
\frac{b_n}{a_n{\!}^2} = \frac{d_n}{2(d_n+2)} \, .
\ee
The coefficients in RG equations such as \eqref{Zeq} should not depend on 
the particular critical point, here labelled by $n$, but may depend on the 
spatial dimension $d$. Applying the $\vep$-expansion to \eqref{Zeq} 
with the particular coefficient $\frac{2d}{d+2}$ ensures, 
as was shown in \cite{DO}  and also subsequently here, that $\eta$ is correct
to ${\rm O}(\vep^2)$ for all critical points $n=2,3,\dots$
Although results such as these for $\eta$ were obtained
from Wilsonian RG equations as soon as they were first proposed,
and were shown to be independent of the detailed cut off function
\cite{Gol}, they are also
identical, of course, with results from standard Feynman graph techniques
which arise directly from the coefficients of the universal logarithmic 
divergencies for particular two point Feynman graphs. These logarithmic
divergencies  are equivalent to \eqref{Feyn}. 
In a sense compatibility with the $\vep$-expansion may be 
regarded as an optimal choice for such constants as $B$. However, 
in the scaling field derivation described in \cite{DO} there is no free constant
to determine and agreement with the $\vep$-expansion is not imposed but
follows automatically. 

In addition \eqref{Zeq} differs from
corresponding equations in \cite{Ball} and \cite{DO} by the absence of a
$\eta Z$ term. Removing such a contribution is essential to obtain subsequent 
results. In general in a derivative expansion there are also expected to be 
additional contributions on the right hand side of
\eqref{LPA} involving $Z$ but the exact form differs between \cite{Ball} and 
\cite{DO} and also involves a cut off dependent constant which we are here 
essentially setting to zero.

Corresponding to \eqref{LPA} and \eqref{Zeq} there are associated eigenvalue
equations for critical exponents
\be
\begin{pmatrix}\D + d & 0 \\
\noalign{\vskip 2pt}
-2Z'(\phi) \frac{\rmd}{\rmd \phi} - 
\big ( 2Z(\phi) - \frac{4d}{d+2}V''(\phi) \big ) \frac{\rmd^2}{\rmd \phi^2} 
& \D - 2 V''(\phi)\end{pmatrix} \begin{pmatrix}f(\phi) \\ g(\phi)\end{pmatrix}  = 
\lambda \begin{pmatrix}f(\phi) \\ g(\phi)\end{pmatrix} \, .
\ee
It is easy to see that this decouples into pairs of equations for eigenvalues
$\lambda_f, \lambda_g$ where $\lambda_f$ is obtained from
\eqref{eig}, with the corresponding $g$ determined in terms of
$f$ by inverting $\D-2V''(\phi)-\lambda_f$, and also, with $f=0$,
\be
\D g (\phi) - 2 V''(\phi) g(\phi) =  \lambda_g \, g(\phi) \, .
\label{eigg}
\ee
For the eigenfunctions in \eqref{exact} the corresponding functions $g$
are given by
\be
- g_\phi (\phi) = g_R(\phi) = Z'(\phi) \, .
\ee

To show reparameterisation invariance of \eqref{Zeq} we note that for any 
solution of \eqref{eig} there is a corresponding solution of \eqref{eigg} 
given by
\be
g(\phi) = f'(\phi) \, , \qquad \lambda_g = \lambda_f - \half (d+2-\eta) \, ,
\ee
in accord with \eqref{red}.
Starting from $f_\phi(\phi)$ in \eqref{exact} it is then easy to obtain an
exact zero mode
\be
g_0(\phi) = \half (2-\eta) - V''(\phi) \, , 
\ee
representing the necessary marginal redundant operator present in the RG flow
equations.  Since $\D - 2 V''(\phi)$ is
hermitian with respect to the scalar product in \eqref{prod} we must have,
for consistent solutions of \eqref{Zeq},
\begin{align}
 \eta \, \langle g_0 , 1 \rangle &{} = \eta \int \rmd \phi \, 
e^{- \frac{1}{4}(d-2+\eta)\phi^2 - 2V(\phi)} \,
\big ( \half (2-\eta) - V''(\phi) \big ) \nn \\
& =  \frac{2d}{d+2} \, \big \langle g_0 , V''^2 \big \rangle +
\big \langle g_0 , ( \D - 2V'') Z \big \rangle \nn \\
& = \frac{2d}{d+2}
\int \rmd \phi \, e^{- \frac{1}{4}(d-2+\eta)\phi^2 - 2V(\phi)} \,
\big ( \half (2-\eta) - V''(\phi) \big ) V''(\phi)^2 \, .
\label{Deta}
\end{align}
Since $\eta$ is small it is easy to iterate \eqref{Deta} in conjunction with
\eqref{LPA} starting from $\eta=0$ to determine the consistent solution for
$\eta$ with high numerical precision.

When $d=3$ we may readily solve \eqref{LPA} numerically tuning $k=V(0)$ so
that the singularity in the solution arises for the largest possible value of 
$\phi$ compatible with numerical  precision,  \eqref{LPA}  was written
as two coupled first order equations and were integrated from $\phi=0$
using RK4.
The limiting results when $\eta=0$
are shown in Figure 1.
\begin{figure}[h]
\begin{center}
\includegraphics[trim= 20mm 80mm 20mm 70mm, clip, scale=0.5]{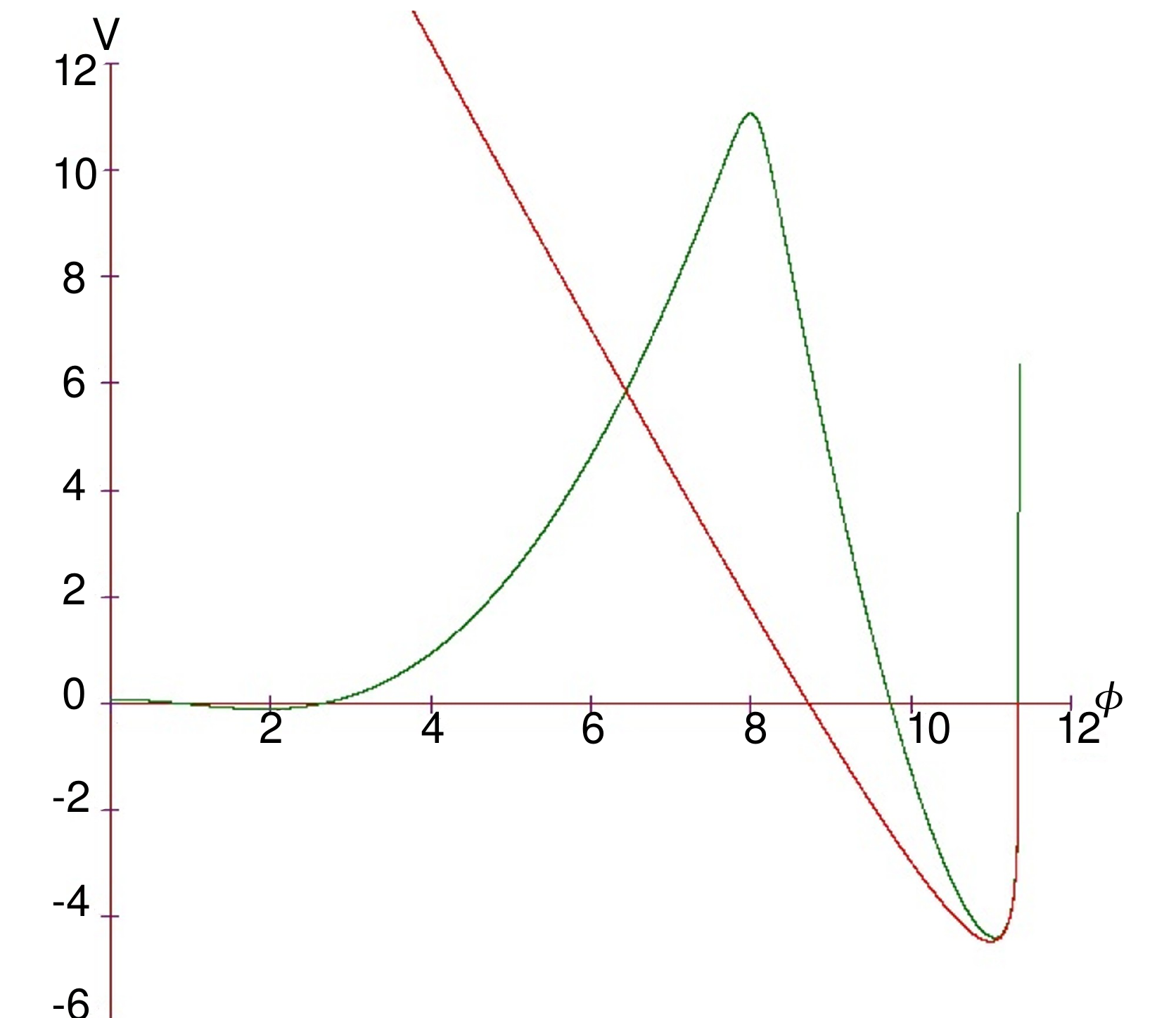}
\end{center}
\caption
{Typical numerical solution of \eqref{LPA}, with $d=3,\eta=0$, for $V(\phi)$ 
starting from the critical value $k=V(0)=0.0761994008$. Due to rounding errors 
the solution breaks  down for $\phi \gtrsim 7$ and is singular at 
$\phi_0\approx 11.23$. As shown, for $\phi \lessapprox \phi_0$,  it is well
approximated near the singularity by the leading singular form 
$V_{\rm sing.}(\phi) = -  \ln ( \phi_0 - \phi) + \quar\phi_0(\phi_0-\phi) + \mbox{const.}$ 
for solutions of \eqref{LPA}. Note that this has a minimum at $\phi=\phi_0-4/\phi_0$
matching the minimum of the numerical solution.
}
\end{figure}
Numerical results for the $Z$-equation \eqref{Zeq} are also shown in Figure 2.
These were obtained in a similar fashion as for $V$ in terms of corresponding
first order differential equations.
The solutions also develop singularities which are very sensitive to the
value of $\eta$, where the corresponding $V$-solution of course has been used.
\begin{figure}[h!]
\centering
\subfloat[eta=0.041346]{\includegraphics[trim= 20mm 70mm 20mm 70mm, clip,
scale=0.45]{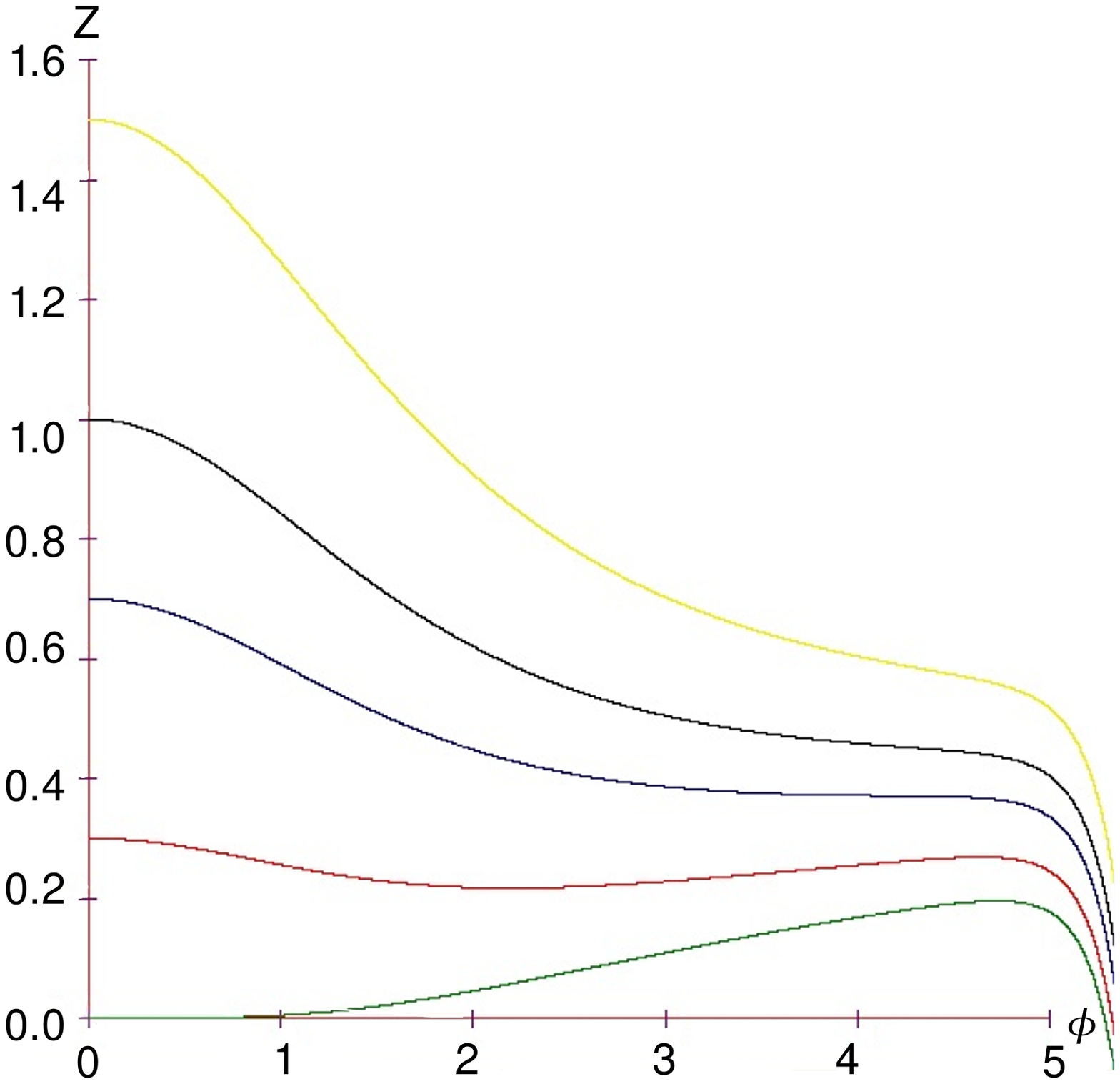}}
\subfloat[eta=0.041347]{\includegraphics[trim= 20mm 70mm 20mm 70mm, clip,
scale=0.45]{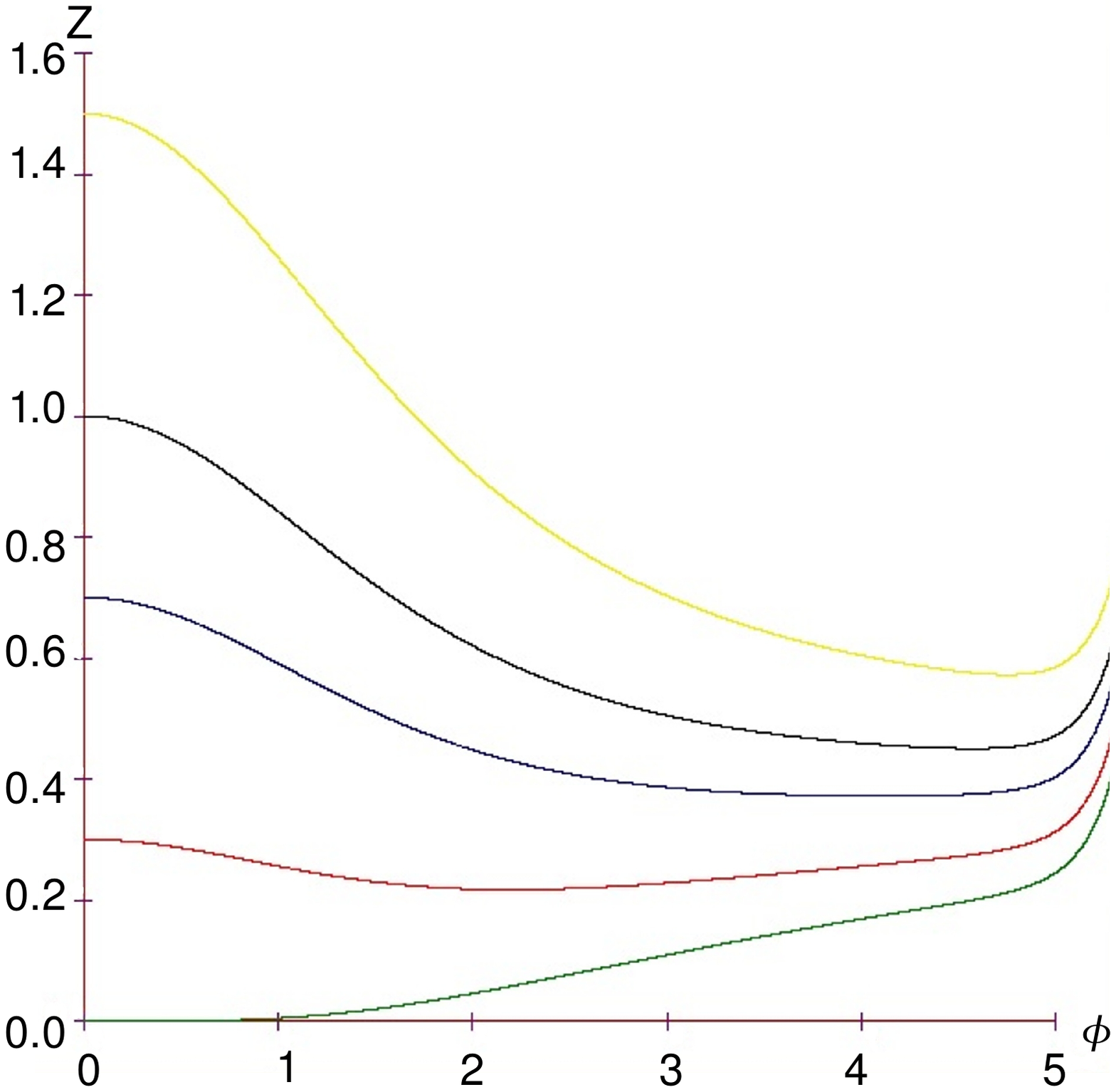}}
\caption{Numerical solutions of \eqref{Zeq} for various $Z(0)$ with $\eta$
just either side of the critical value so that the singularity arises
for the largest possible $\phi$. The graphs demonstrate how $\eta$ is 
independent of $Z(0)$.}
\end{figure}\\

Having determined $V$ the eigenvalues $\lambda_n$ are then determined 
numerically for small values of $\eta$ and $d=3$ by optimising
the eigenvalue such that the eigenfunction blows up as slowly as possible
withing the range where results for $V(\phi)$ are reliable. 
The results are ordered such 
that $\lambda_n > \lambda_{n+1}$ with $\lambda_0=3$. For $n$  even,odd the 
associated eigenfunctions are even,odd in $\phi$. From \eqref{exact}
$\lambda_1 = \half (5-\eta)$ and $\lambda_3 = \half (1+\eta)$ which provides
a consistency check on our numerical results. For even $n$ the results are in 
Table 1 and for odd $n$ in Table 2. For the small values considered the 
dependence on $\eta$ is close to linear. For $\eta=0$ our results agree with 
the much more accurate determinations in \cite{Acc,Con}.

\begin{table}[!h]
\begin{tabular}{|r|r|r|r|r|r|r|} \hline
$\eta$~~&$k\qquad~~$&$\lambda_2\quad$&$\lambda_4\quad$&$\lambda_6\quad$
&$\lambda_8\quad$&$\lambda_{10}\quad$\\
\hline
0.00&\hfil$0.076199401$~~&\hfil$1.5395$~~&\hfil$-0.6557$~~&\hfil$-3.1800$
~~&\hfil$-5.9122$~~&\hfil$-8.7961$\\
0.01&\hfil$0.073512228$~~&\hfil$1.5409$~~&\hfil$-0.6449$~~&\hfil$-3.1560$
~~&\hfil$-5.8735$~~&\hfil$-8.7416$\\
0.02&\hfil$0.070885912$~~&\hfil$1.5421$~~&\hfil$-0.6341$~~&\hfil$-3.1319$
~~&\hfil$-5.8344$~~&\hfil$-8.6866$\\
0.03&\hfil$0.068319137$~~&\hfil$1.5433$~~&\hfil$-0.6232$~~&\hfil$-3.1076$
~~&\hfil$-5.7951$~~&\hfil$-8.6312$\\
0.04&\hfil$0.065810847$~~&\hfil$1.5444$~~&\hfil$-0.6123$~~&\hfil$-3.0832$
~~&\hfil$-5.7554$~~&\hfil$-8.5754$\\
0.05&\hfil$0.063359963$~~&\hfil$1.5454$~~&\hfil$-0.6013$~~&\hfil$-3.0585$
~~&\hfil$-5.7155$~~&\hfil$-8.5190$\\
0.06&\hfil$0.060965439$~~&\hfil$1.5463$~~&\hfil$-0.5903$~~&\hfil$-3.0337$
~~&\hfil$-5.6751$~~&\hfil$-8.4621$\\
0.07&\hfil$0.058626258$~~&\hfil$1.5471$~~&\hfil$-0.5793$~~&\hfil$-3.0087$
~~&\hfil$-5.6345$~~&\hfil$-8.4048$\\
0.08&\hfil$0.056341435$~~&\hfil$1.5478$~~&\hfil$-0.5682$~~&\hfil$-2.9835$
~~&\hfil$-5.5935$~~&\hfil$-8.3469$\\
0.09&\hfil$0.054110012$~~&\hfil$1.5485$~~&\hfil$-0.5570$~~&\hfil$-2.9582$
~~&\hfil$-5.5521$~~&\hfil$-8.2884$\\
0.10&\hfil$0.051931059$~~&\hfil$1.5490$~~&\hfil$-0.5458$~~&\hfil$-2.9326$
~~&\hfil$-5.5103$~~&\hfil$-8.2294$\\
\hline
\end{tabular}\caption{Even Eigenvalues and initial value $k=V(0)$ for
non singular solutions}
\end{table}
\begin{table}[!h]
\begin{tabular}{|r|r|r|r|r|} \hline
$\eta$~~&$\lambda_5\quad$&$\lambda_7\quad$&$\lambda_9\quad$\\
\hline
0.00&\hfil$-1.8867$~~&\hfil$-4.5244$~~&\hfil$-7.3377$\\
0.01&\hfil$-1.8696$~~&\hfil$-4.4932$~~&\hfil$-7.2911$\\
0.02&\hfil$-1.8524$~~&\hfil$-4.4618$~~&\hfil$-7.2422$\\
0.03&\hfil$-1.8351$~~&\hfil$-4.4301$~~&\hfil$-7.1970$\\
0.04&\hfil$-1.8177$~~&\hfil$-4.3982$~~&\hfil$-7.1493$\\
0.05&\hfil$-1.8002$~~&\hfil$-4.3660$~~&\hfil$-7.1012$\\
0.06&\hfil$-1.7826$~~&\hfil$-4.3336$~~&\hfil$-7.0527$\\
0.07&\hfil$-1.7648$~~&\hfil$-4.3010$~~&\hfil$-7.0038$\\
0.08&\hfil$-1.7470$~~&\hfil$-4.2680$~~&\hfil$-6.9545$\\
0.09&\hfil$-1.7290$~~&\hfil$-4.2348$~~&\hfil$-6.9047$\\
0.10&\hfil$-1.7109$~~&\hfil$-4.2013$~~&\hfil$-6.8544$\\
\hline
\end{tabular}\caption{Odd Eigenvalues}
\end{table}

We have also determined the value of $\eta$ required for non
singular solutions of \eqref{Zeq} and verified that the result is
independent of the value chosen for $Z(0)$ and also in agreement
with \eqref{Deta}. This may be used to determine $\eta$ by iteration
starting from $\eta=0$ and gives
\be
\eta =   0.041347    \qquad \mbox{for}  \qquad  k = 0.0654776 \, .
\ee
For this value
\be
\nu = 1/\lambda_2 = 0.647 \, , \qquad \omega = - \lambda_4 = 0.612 \, .
\ee

\section{Multi-Component Fields}

There are natural generalisations of the above equations to the 
case of a $N$-component scalar field $\phi_i$. Instead of \eqref{LPA}
we have
\be
\pr^2 V(\phi) -\half \big ( (d-2)\phi + (\eta\phi)\big  )\cdot 
\frac{\pr}{\pr \phi} V(\phi) - \frac{\pr}{\pr \phi} V(\phi)
\cdot \frac{\pr}{\pr \phi} V(\phi) + dV(\phi)  =0 \, ,
\label{LPA2}
\ee
for $\eta_{ij}$ a symmetric anomalous dimension matrix and
$\pr^2 = \frac{\pr}{\pr \phi} \cdot \frac{\pr}{\pr \phi}$.
The corresponding
equation for critical exponents is just as in \eqref{eig},
\be
\D F(\phi) = (\lambda - d) F(\phi) \, ,
\label{eigF}
\ee
but with the differential operator        
\be
\D = \pr^2 -\half \big ( (d-2)\phi + (\eta\phi)\big  )\cdot \frac{\pr}{\pr \phi} 
- 2 \frac{\pr}{\pr \phi} V(\phi) \cdot \frac{\pr}{\pr \phi} \, ,
\label{diff2}
\ee
and we may now also allow tensorial eigenfunctions $F_{ij\dots}(\phi)$. 
Trivially $\D 1 =0$ so that $\lambda=d$ is an exact eigenvalue.
Just as in \eqref{exact} we have exact vector eigenfunctions since
\be
\D F_{\phi,i}(\phi) = \half(d-2) F_{\phi,i}(\phi) + \half \big ( 
\eta F_{\phi}(\phi) \big ) {}_i \, , \qquad F_{\phi,i}(\phi) = \phi_i
- \half (\eta \phi)_i - \frac{\pr}{\pr \phi_i} V(\phi) \, ,
\label{fphi}
\ee
and we may choose a diagonal basis for $\eta_{ij}$.
The corresponding scalar product to \eqref{prod} is just
\be
\langle F ,G \rangle = \int \rmd^N \! \phi \; e^{- \frac{1}{4}(d-2)\phi^2
- \frac{1}{4} \phi \cdot (\eta \phi) - 2V(\phi)} \, F(\phi)\, G(\phi) \, ,
\label{prod2}
\ee
with additional tensorial contractions if required.

Extending \eqref{Zeq} to $Z_{ij}(\phi)$ we require
\be
\D Z_{ij}(\phi) - 2 V_{(ik}(\phi) Z_{j)k}(\phi) = \eta_{ij}
- \frac{2d}{d+2} \, V_{ik}(\phi) V_{jk}(\phi) \, , \qquad
V_{ij}(\phi) = \frac{\pr^2}{\pr \phi_i \pr \phi_j} V(\phi) \, ,
\label{Z2}
\ee
with $( \  )$ used to denote symmetrisation of the $i,j$ indices.
For fluctuations $F,G_{ij}$ around the fixed point solutions we have \eqref{eigF}
and also the associated coupled equation
\begin{align}
& \D G_{ij} (\phi) - 2 V_{(ik}(\phi) G_{j)k}(\phi) \nn \\
& - 2 \pr_k Z_{ij}(\phi)\, \pr_k F(\phi) 
- \bigg ( 2Z_{(ik}(\phi) - \frac{4d}{d+2}V_{(ik}(\phi) \bigg )\pr_{j)} \pr_k F(\phi)
=  \lambda G_{ij}(\phi) \, .
\end{align}
As before it is easy to see that the possible eigenvalues are $\lambda_F$
determined by  \eqref{eigF} and $\lambda_G$ given by
\be
\D G_{ij} (\phi) - 2 V_{(ik}(\phi) G_{j)k}(\phi) =  \lambda_G G_{ij}(\phi) \, ,
\label{eigg2}
\ee
corresponding to \eqref{eigg}.
For any vector eigenfunction $F_i(\phi)$ there are associated eigenfunctions
for \eqref{eigg2} given in terms of 
\be
G_{ij}(\phi) = \pr_i F_j(\phi) + \pr_j F_i(\phi) \, , 
\ee
since
\be
\D G_{ij} (\phi) - 2 V_{(ik}(\phi) G_{j)k}(\phi) =  \big ( \lambda_F
- \half (d-2) \big )  G_{ij}(\phi) - \half  \eta_{(ik} G_{j)k}(\phi)\, ,
\label{eiggz}
\ee
Hence we have an exact zero mode which may be obtained from \eqref{fphi}
\be
G_{0,ij}(\phi) = \delta_{ij} - \half \eta_{ij} - V_{ij}(\phi) \, .
\ee
In consequence for \eqref{Z2} to be solvable we must have
\be
\int \rmd^N \! \phi \; e^{- \frac{1}{4}(d-2)\phi^2
- \frac{1}{4} \phi \cdot (\eta \phi) - 2V(\phi)} \,
\big ( \delta_{ij} - \half \eta_{ij} - V_{ij}(\phi)  \big )
\bigg ( \eta_{ij} - \frac{2d}{d+2} \, V_{ik}(\phi) V_{jk}(\phi) \bigg ) =0 \, .
\label{etg}
\ee

In order to obtain tractable equations we impose $O(N)$ symmetry so that
we need only deal with functions of $\rho = \half \phi^2$ and in this
case we must have $\eta_{ij}= \eta \delta_{ij}$. Writing $V(\phi)=v(\rho)$
\eqref{LPA2} becomes
\be
2\rho v''(\rho) - (d-2+\eta) \rho v'(\rho) = - Nv'(\rho)+2\rho v'(\rho)^2
- d v(\rho) \, .
\label{LPA3}
\ee
At the origin we must have $Nv'(0)+dv(0)=0$ and asymptotically
\be
v(\rho) \sim \half(2-\eta) \rho + A \, \rho^{\frac{d}{d+2-\eta}} \, .
\ee
To ensure non singular solutions as before it is necessary to fine tune $v(0)$.
For critical exponents we consider, if $N>1$, spherical harmonics which are 
expressible in terms of symmetric traceless tensors or equivalently
\be
F_l (\phi) = (t\cdot \phi)^l \, f_l (\rho) \, , \quad t^2 =0 \, , \qquad
l=0,1,2,\dots \, .
\ee
The eigenvalue equation \eqref{eigF} becomes
\be
\D_l f_l (\rho) = \big ( \lambda_l + l \half( d - 2 + \eta) -d \big ) 
f_l (\rho) \, ,
\label{eigf}
\ee
for
\be 
\D_l = 2\rho \frac{\rmd^2}{\rmd \rho^2} - (d-2+\eta) \rho \frac{\rmd}{\rmd\rho}
+ (N+2l) \frac{\rmd}{\rmd\rho} - 4\rho v'(\rho) \frac{\rmd}{\rmd\rho} 
- 2l v'(\rho) \, .
\label{Dop}
\ee
The relevant boundary conditions are that $f_l (\rho)$ is analytic for 
$\rho\approx 0$ and non singular for $\rho>0$ and $\propto 
\rho^{\frac{d-\lambda_l}{d+2-\eta}-\frac{1}{2} l}$ as $\rho \to \infty$.
Corresponding to \eqref{fphi} 
\be
f_{1,\phi}(\rho) = \half (2-\eta) - v'(\rho) \, ,
\label{zf}
\ee
satisfies
\be
\D_1 f_{1,\phi}(\rho) = 0 \, ,
\label{zzf}
\ee
so that $\lambda_{1,\phi} = \half (d+2-\eta)$.
The scalar product, with respect to which $\D_l$ is hermitian, becomes
from \eqref{prod2}
\be
\langle f_l , g_l \rangle_l = \int_0^\infty \!\!\! \rmd \rho  \; 
\rho^{\frac{1}{2}N + l -1}
e^{- \frac{1}{2}(d-2+\eta)\rho - 2v(\rho)} \, f_l(\rho)\, g_l(\rho) \, ,
\label{prod3}
\ee
When $N=1$ only $l=0,1$ are relevant, corresponding to even,odd eigenfunctions.
In terms of $v(\rho)$, which solves \eqref{LPA3}, \eqref{etg} becomes
\begin{align}
& \eta \int_0^\infty \!\!\! \rmd \rho  \;
\rho^{\frac{1}{2}N -1} e^{- \frac{1}{2}(d-2+\eta)\rho - 2v(\rho)} \,
\big ( \half (2-\eta)N - Nv'(\rho) - 2\rho v''(\rho) \big ) \nn \\
& = \frac{2d}{d+2} \int_0^\infty \!\!\! \rmd \rho  \;
\rho^{\frac{1}{2}N  -1} e^{- \frac{1}{2}(d-2+\eta)\rho - 2v(\rho)} \nn \\
\noalign{\vskip -4pt}
& \hskip 2cm {}\times \Big ( \half (2-\eta)
\big ( N v'(\rho)^2 + 4 \rho v'(\rho) v''(\rho) + 4 \rho^2 v''(\rho)^2 \big ) \nn \\
\noalign{\vskip -4pt}
& \hskip 2.7cm {} - N v'(\rho)^3 - 6 \rho v'(\rho)^2 v''(\rho) -12
\rho^2 v'(\rho) v''(\rho)^2 - 8 \rho^3 v''(\rho)^3 \Big ) \, .
\label{deta}
\end{align}
When $N=1$ it is easy to see that this is identical with \eqref{Deta} since
then $V''(\phi) = v'(\rho) + 2\rho v'(\rho)$.

To decompose \eqref{Z2} we write
\be
Z_{ij}(\phi) = \delta_{ij}\, z(\rho) + \phi_i \phi_j \big ( z'(\rho) + y(\rho) \big )
= \pr_{(i}\big ( \phi_{j)} z(\rho) \big ) + \phi_i \phi_j y(\rho) 
\, ,
\label{Zdecom}
\ee
and using
\begin{align}
\D Z_{ij} (\phi) - 2 V_{(ik}(\phi) Z_{j)k}(\phi) 
={} &\delta_{ij} \big ( \D_1 z(\rho) + 2 y(\rho) \big ) 
+ \phi_i \phi_j \, \frac{\rmd}{\rmd \rho} \big ( \D_1 z(\rho) + 2 y(\rho) \big )\nn\\
&{}+ \phi_i \phi_j \big ( \D_1 y(\rho) - (d-2+\eta + 4 v'(\rho)+4\rho v''(\rho))
y(\rho) \big ) \, ,
\end{align}
we may reduce \eqref{Z2} to
\be
\D_1 z(\rho) = \eta - 2 y(\rho) - \frac{2d}{d+2} \, v'(\rho)^2 \, ,
\label{zeq}
\ee
and
\be
\D_1 y(\rho) - \big ( d-2 + \eta + 4 \rho v''(\rho) + 4 v'(\rho) \big ) y(\rho)
= - \frac{4d}{d+2} \, \rho v''(\rho)^2 \, .
\label{yeq}
\ee
The equation for $y$ thus decouples from that for $z$ so that 
\eqref{yeq} may be solved and  the result used in \eqref{zeq} to then
determine $z$. Asymptotically $z(\rho)$ approaches a constant, just as in 
\eqref{asymZ}, but $y(\rho)$ vanishes.
Since, as shown in \eqref{zzf}, $\D_1$ has a zero mode 
$f_{1,\phi}$, given by \eqref{zf}, $\eta$
must be fixed to allow a non singular solution by,
\be
\eta\, \big \langle f_{1,\phi} , 1 \big \rangle_1
= 2 \, \big \langle f_{1,\phi} , y \big \rangle_1 + \frac{2d}{d+2} \,
\big \langle f_{1,\phi} , v'^2  \big \rangle_1 \, .
\label{e1}
\ee
Using 
\begin{align}
\int_0^\infty \!\!\! \rmd \rho  \;
\rho^{\frac{1}{2}N -1}&  e^{- \frac{1}{2}(d-2+\eta)\rho - 2v(\rho)} \, 
G_{0,ij}(\phi) \phi_i \phi_j \nn \\
\noalign{\vskip -4pt}
& {}\times \big ( 
\D_1 y(\rho) - ( d-2 + \eta + 4 \rho v''(\rho) + 4 v'(\rho) ) y(\rho)\big )\nn \\
&=  - 2 (d-2+\eta) \, \big \langle f_{1,\phi} , y \big \rangle_1 
\end{align}
we may also obtain
\be
(d-2+\eta) \, \big \langle f_{1,\phi} , y \big \rangle_1 
= \frac{4d}{d+2} \int_0^\infty \!\!\! \rmd \rho  \;
\rho^{\frac{1}{2}N +1} e^{- \frac{1}{2}(d-2+\eta)\rho - 2v(\rho)} \,
\big ( \half (2-\eta) - v'(\rho) - 2\rho v''(\rho) \big ) v''(\rho)^2 \, .
\label{e2}
\ee
Combining \eqref{e1} and \eqref{e2} is equivalent to \eqref{deta}. If 
we restrict to scalar fluctuations, without any harmonics, then we
may decompose $G_{ij}(\phi)$ in terms of $g(\rho),h(\rho)$ in a similar
fashion to \eqref{Zdecom} so that \eqref{eigg2} reduces to
\be
\D_1 g(\rho) = \lambda g(\rho) \, , 
\label{Dg}
\ee
with $h=0$ and also
\be
\D_1 h(\rho) - 4 \big ( \rho v''(\rho) +  v'(\rho) \big ) h(\rho) =
(\lambda +  d-2 + \eta )  h(\rho) \, ,
\label{Dh}
\ee
when $(\D_1 - \lambda) g(\rho)=-2h(\rho)$.
Manifestly the eigensolutions for $g$ in \eqref{Dg}
are identical with the solutions of \eqref{eigf} for $l=1$. 
The eigenvalues are related as expected from
\eqref{red} so the solutions of \eqref{Dg} represent redundant operators. 
The eigenvalues determined by the $h$-equation \eqref{Dh} give exponents 
corresponding to genuine physical scaling operators.

Numerically \eqref{LPA3} and \eqref{eigf} can be solved straightforwardly for 
$d=3$, as before after precisely tuning $k^{(N)}= v(0)$ to ensure non singular 
solutions for all $\rho >0$. The eigenvalues are denoted by 
$\lambda^{(N)}_{l,m}$ where $\lambda^{(N)}_{0,0}=3$ and  we take
$m=0,1,2,\dots$. In terms of the single
component results $\lambda^{(1)}_{0,m} = \lambda_{2m}$ and 
$\lambda^{(1)}_{1,m} = \lambda_{2m+1}$. When $l=0$ some results are given in 
table 3, for $N=1$ they match as expected the results given earlier. 
For $\eta=0$ these agree with results in \cite{Com2,Litim}. In
the large $N$ limit, taking $\eta=0$, then $\lambda^{(N)}_{l,m} = 3-2m -\half l$.

\begin{table}[!h]
\begin{tabular}{|r|r|r|r|r|r|r|r|r|} \hline
$\eta$~~&$k^{(1)}\quad$&$\lambda^{(1)}_{0,1}\quad$&$k^{(2)}\quad$&
$\lambda^{(2)}_{0,1}\quad$&$k^{(4)}\quad$&$\lambda^{(4)}_{0,1}\quad$
&$k^{(10)}\quad$&$\lambda^{(10)}_{0,1}\quad$\\
\hline
0.00&\hfil$0.076200$~&\hfil$1.5395$~&\hfil$0.19934$~~&\hfil$1.4120$
~&\hfil$0.53876$~&\hfil$1.2432$~&\hfil$1.7637$~&\hfil$1.0886$\\
0.01&\hfil$0.073513$~&\hfil$1.5409$~&\hfil$0.19176$~~&\hfil$1.4175$
~&\hfil$0.51693$~&\hfil$1.2535$~&\hfil$1.6929$~&\hfil$1.0999$\\
0.02&\hfil$0.070886$~&\hfil$1.5421$~&\hfil$0.18439$~~&\hfil$1.4229$
~&\hfil$0.49578$~&\hfil$1.2634$~&\hfil$1.6242$~&\hfil$1.1112$\\
0.03&\hfil$0.068319$~&\hfil$1.5433$~&\hfil$0.17723$~~&\hfil$1.4281$
~&\hfil$0.47534$~&\hfil$1.2732$~&\hfil$1.5577$~&\hfil$1.1224$\\
0.04&\hfil$0.065811$~&\hfil$1.5444$~&\hfil$0.17026$~~&\hfil$1.4330$
~&\hfil$0.45552$~&\hfil$1.2823$~&\hfil$1.4931$~&\hfil$1.1336$\\
0.05&\hfil$0.063360$~&\hfil$1.5454$~&\hfil$0.16349$~~&\hfil$1.4378$
~&\hfil$0.43633$~&\hfil$1.2921$~&\hfil$1.4306$~&\hfil$1.1447$\\
\hline
\end{tabular}\caption{First $l=0$ Eigenvalues for $N=1,2,4,10$}
\end{table}

\begin{table}[!h]
\begin{tabular}{|r|r|r|r|r|r|r|r|r|} \hline
$\eta$~~&$\lambda^{(1)}_{0,2}\quad$&$\lambda^{(2)}_{0,2}\quad$&
$\lambda^{(4)}_{0,2}\quad$&
$\lambda^{(10)}_{0,2}\quad$&$\lambda^{(1)}_{0,3}\quad$&$\lambda^{(2)}_{0,3}\quad$
&$\lambda^{(4)}_{0,3}\quad$&$\lambda^{(10)}_{0,3}\quad$\\
\hline
0.00&\hfil$-0.6557$~&\hfil$-0.6712$~&\hfil$-0.7338$~~&\hfil$-0.8713$
~&\hfil$-3.1800$~&\hfil$-3.0714$~&\hfil$-2.9400$~&\hfil$-2.8985$\\
0.01&\hfil$-0.6449$~&\hfil$-0.6596$~&\hfil$-0.7187$~~&\hfil$-0.8515$
~&\hfil$-3.1560$~&\hfil$-3.0498$~&\hfil$-2.9199$~&\hfil$-2.8728$\\
0.02&\hfil$-0.6341$~&\hfil$-0.6480$~&\hfil$-0.7039$~~&\hfil$-0.8319$
~&\hfil$-3.1319$~&\hfil$-3.0281$~&\hfil$-2.8998$~&\hfil$-2.8473$\\
0.03&\hfil$-0.6232$~&\hfil$-0.6364$~&\hfil$-0.6892$~~&\hfil$-0.8124$
~&\hfil$-3.1076$~&\hfil$-3.0062$~&\hfil$-2.8795$~&\hfil$-2.8221$\\
0.04&\hfil$-0.6123$~&\hfil$-0.6247$~&\hfil$-0.6746$~~&\hfil$-0.7930$
~&\hfil$-3.0832$~&\hfil$-2.9841$~&\hfil$-2.8592$~&\hfil$-2.7970$\\
0.05&\hfil$-0.6013$~&\hfil$-0.6131$~&\hfil$-0.6602$~~&\hfil$-0.7739$
~&\hfil$-3.0585$~&\hfil$-2.9617$~&\hfil$-2.8397$~&\hfil$-2.7722$\\
\hline
\end{tabular}\caption{Second and Third $l=0$ Eigenvalues for $N=1,2,4,10$}
\end{table}

For $l=1$ then $\lambda^{(N)}_{1,0}= \half (5-\eta)$ and
$\lambda^{(N)}_{1,1}= \half (1+\eta)$ which is a useful check. Some other
results are given in table 5. We also present some results for $l=2$ in 
table 6 and $l=4$ in table 7. An important observation is that
$\lambda^{(2)}_{4,0}<0$ whereas $\lambda^{(N)}_{4,0}> 0$ for $N \ge 3$.
This reflects the instability of the $O(N)$ symmetric fixed point against
RG flow to one with just discrete cubic symmetry when $N\ge 3$. That the critical
$N_c<3$ has been shown in very detailed multi-loop Feynman diagram calculations
\cite{Hagen}.

\begin{table}[!h]
\begin{tabular}{|r|r|r|r|r|r|r|r|r|} \hline
$\eta$~~&$\lambda^{(1)}_{1,2}\quad$&$\lambda^{(2)}_{1,2}\quad$&
$\lambda^{(4)}_{1,2}\quad$&
$\lambda^{(10)}_{1,2}\quad$&$\lambda^{(1)}_{1,3}\quad$&$\lambda^{(2)}_{1,3}\quad$
&$\lambda^{(4)}_{1,3}\quad$&$\lambda^{(10)}_{1,3}\quad$\\
\hline
0.00&\hfil$-1.8867$~&\hfil$-1.7986$~&\hfil$-1.6741$~~&\hfil$-1.5535$
~&\hfil$-4.5244$~&\hfil$-4.3251$~&\hfil$-4.0185$~&\hfil$-3.6719$\\
0.01&\hfil$-1.8696$~&\hfil$-1.7835$~&\hfil$-1.6615$~~&\hfil$-1.5406$
~&\hfil$-4.4932$~&\hfil$-4.2984$~&\hfil$-3.9983$~&\hfil$-3.6529$\\
0.02&\hfil$-1.8524$~&\hfil$-1.7683$~&\hfil$-1.6487$~~&\hfil$-1.5277$
~&\hfil$-4.4618$~&\hfil$-4.2714$~&\hfil$-3.9778$~&\hfil$-3.6339$\\
0.03&\hfil$-1.8351$~&\hfil$-1.7530$~&\hfil$-1.6358$~~&\hfil$-1.5148$
~&\hfil$-4.4301$~&\hfil$-4.2442$~&\hfil$-3.9570$~&\hfil$-3.6149$\\
0.04&\hfil$-1.8177$~&\hfil$-1.7375$~&\hfil$-1.6228$~~&\hfil$-1.5020$
~&\hfil$-4.3982$~&\hfil$-4.2166$~&\hfil$-3.9358$~&\hfil$-3.5958$\\
0.05&\hfil$-1.8002$~&\hfil$-1.7220$~&\hfil$-1.6096$~~&\hfil$-1.4891$
~&\hfil$-4.3661$~&\hfil$-4.1887$~&\hfil$-3.9143$~&\hfil$-3.5767$\\
\hline
\end{tabular}\caption{$l=1$ Eigenvalues for $N=1,2,4,10$}
\end{table}

\begin{table}[!h]
\begin{tabular}{|r|r|r|r|r|r|r|} \hline
$\eta$~~&$\lambda^{(2)}_{2,0}\quad$&$\lambda^{(4)}_{2,0}\quad$&
$\lambda^{(10)}_{2,0}\quad$&
$\lambda^{(2)}_{2,1}\quad$&$\lambda^{(4)}_{2,1}\quad$&$\lambda^{(10)}_{2,1}\quad$\\
\hline
0.00&\hfil$1.7819$~&\hfil$1.8476$~&\hfil$1.9283$~~&\hfil$-0.4737$
~&\hfil$-0.3332$~&\hfil$-0.1531$\\
0.01&\hfil$1.7760$~&\hfil$1.8396$~&\hfil$1.9187$~~&\hfil$-0.4675$
~&\hfil$-0.3308$~&\hfil$-0.1535$\\
0.02&\hfil$1.7701$~&\hfil$1.8317$~&\hfil$1.9091$~~&\hfil$-0.4612$
~&\hfil$-0.3282$~&\hfil$-0.1539$\\
0.03&\hfil$1.7642$~&\hfil$1.8239$~&\hfil$1.8996$~~&\hfil$-0.4548$
~&\hfil$-0.3255$~&\hfil$-0.1541$\\
0.04&\hfil$1.7583$~&\hfil$1.8161$~&\hfil$1.8901$~~&\hfil$-0.4482$
~&\hfil$-0.3226$~&\hfil$-0.1542$\\
0.05&\hfil$1.7525$~&\hfil$1.8084$~&\hfil$1.8807$~~&\hfil$-0.4416$
~&\hfil$-0.3195$~&\hfil$-0.1543$\\
\hline
\end{tabular}\caption{$l=2$ Eigenvalues for $N=2,4,10$}
\end{table}

\begin{table}[!h]
\begin{tabular}{|r|r|r|r|r|r|r|r|r|} \hline
$\eta$~~&$\lambda^{(2)}_{4,0}\quad$&$\lambda^{(3)}_{4,0}\quad$&
$\lambda^{(4)}_{4,0}\quad$&
$\lambda^{(10)}_{4,0}\quad$&$\lambda^{(2)}_{4,1}\quad$&$\lambda^{(3)}_{4,1}\quad$
&$\lambda^{(4)}_{4,1}\quad$&$\lambda^{(10)}_{4,1}\quad$\\
\hline
0.00&\hfil$-0.0337$~&\hfil$0.1109$~&\hfil$0.2315$~~&\hfil$0.6045$
~&\hfil$-2.6147$~&\hfil$-2.3954$~&\hfil$-2.2093$~&\hfil$-1.6169$\\
0.01&\hfil$-0.0358$~&\hfil$0.1046$~&\hfil$0.2218$~~&\hfil$0.5871$
~&\hfil$-2.6023$~&\hfil$-2.3889$~&\hfil$-2.2077$~&\hfil$-1.6286$\\
0.02&\hfil$-0.0377$~&\hfil$0.0985$~&\hfil$0.2124$~~&\hfil$0.5700$
~&\hfil$-2.5897$~&\hfil$-2.3821$~&\hfil$-2.2056$~&\hfil$-1.6373$\\
0.03&\hfil$-0.0395$~&\hfil$0.0926$~&\hfil$0.2032$~~&\hfil$0.5530$
~&\hfil$-2.5768$~&\hfil$-2.3749$~&\hfil$-2.2031$~&\hfil$-1.6456$\\
0.04&\hfil$-0.0412$~&\hfil$0.0869$~&\hfil$0.1943$~~&\hfil$0.5326$
~&\hfil$-2.5636$~&\hfil$-2.3673$~&\hfil$-2.2001$~&\hfil$-1.6536$\\
0.05&\hfil$-0.0428$~&\hfil$0.0815$~&\hfil$0.1856$~~&\hfil$0.5197$
~&\hfil$-2.5501$~&\hfil$-2.3593$~&\hfil$-2.1967$~&\hfil$-1.6612$\\
\hline
\end{tabular}\caption{$l=4$ Eigenvalues for $N=2,3,4,10$}
\end{table}

We may also use \eqref{deta} and also \eqref{e1} with \eqref{e2} to
determine $\eta^{(N)}$ when $d=3$ with the results
\begin{align}
\eta^{(1)} = {}& 0.0413 \, , \quad \eta^{(2)} = 0.0414 \, , \quad
\eta^{(3)} = 0.0390 \, , \quad \eta^{(4)} = 0.0357 \, , \nn \\
\eta^{(10)} = {}& 0.0200 \, , \qquad \eta^{(20)} = 0.0125 \, ,
\end{align}
falling off as expected with increasing $N$.

In order to verify consistency with large $N$ results we also calculated
eigenvalues for $N=20$ and $\eta=0$, when $k^{(20)}=3.8727448$, 
which are given in table 8. For $l,m$ small these are closed to the
asymptotic values.
\begin{table}[!h]
\begin{tabular}{|r|r|r|r|r|r|r|} \hline
$m=$~~&$0\quad$&$1\quad$&$2\quad$&
$3\quad$&$4\quad$&$5\quad$\\
\hline
$\lambda^{(20)}_{0,m}$&\hfil$3.000$~&\hfil$1.041$~&\hfil$-0.937$~~&\hfil$-2.938$
~&\hfil$-4.966$~&\hfil$-7.025$\\
$\lambda^{(20)}_{1,m}$&\hfil$2.500$~&\hfil$0.500$~&\hfil$-1.521$~~&\hfil$-3.566$
~&\hfil$-5.640$~&\hfil$-7.744$\\
$\lambda^{(20)}_{2,m}$&\hfil$1.963$~&\hfil$-0.077$~&\hfil$-2.139$~~&\hfil$-4.227$
~&\hfil$~$~&\hfil$~$\\
$\lambda^{(20)}_{3,m}$&\hfil$1.392$~&\hfil$-0.687$~&\hfil$-2.789$~~&\hfil$-4.917$
~&\hfil$~$~&\hfil$~$\\
$\lambda^{(20)}_{4,m}$&\hfil$0.788$~&\hfil$-1.328$~&\hfil$-3.468$~~&\hfil$-5.634$
~&\hfil$~$~&\hfil$~$\\
\hline
\end{tabular}\caption{Eigenvalues for $N=20$}
\end{table}

An interesting special case is $N=-2$, which was considered in \cite{Morris3}.
For this $N$ from \eqref{LPA3} we get $v'(0)(v'(0)-1+\half \eta)=0$. When
$v'(0) = 1-\half \eta$ we have the high temperature fixed point solution
$v(\rho) = (1-\half \eta)(\rho+ \frac{2}{d})$. The relevant non trivial
fixed point arises for $v(0)=v'(0)=0$. Consistency with our
equations requires $\eta^{(-2)}=0$. To show this we may note in 
\eqref{zf} that $f_{1,\phi}(\rho) \to \half(2-\eta)$ as $\rho \to 0$ 
so that,  with the definition of the scalar product in \eqref{prod3},
the integral $\langle f_{1,\phi} , 1 \rangle_1$ diverges due to the singular 
behaviour as $\rho \to 0$. 
On the other hand $\langle f_{1,\phi} , v'^2 \rangle_1$ and also
$\langle f_{1,\phi} , y \rangle_1$, given by \eqref{e2}, are finite.  Hence
\eqref{e1} is only consistent when $\eta=0$. For the corresponding 
eigenfunctions it is necessary to consider boundary conditions carefully. 
In \eqref{eigf} then as $\rho\to 0$ the differential equation requires 
solutions for $f_l(\rho)$ which are ${\rm O}(1)$ and ${\rm O}(\rho^a)$ where 
$a=1-\half(N+2l) -2Nk/d,\, k=v(0)$.
Generally we impose the requirement that the second solution is absent.
For $N=-2, \, l=0$, when also $k=0$, there are solutions with $f_0(\rho)=
{\rm O}(\rho^2)$ for which the norm given by \eqref{prod3} is finite. 
However to obtain eigenvalues which are related to those for $N\ge 0$ it
appears necessary to consider as before solutions with $f_0(0)=1$, and we 
may impose $f_0{}''(0)=0$. In this case we find numerically $\lambda_{0,1}=2$.

\section{$\vep$ Expansion}

A further consistency check, following \cite{DO}, is to consider the 
$\vep$-expansion where
\be
d = \frac{2n}{n-1}-\vep  \, , \qquad \frac{d}{d-2} = n + \half(n-1)^2 \vep
+ {\rm O}(\vep^2) \, , \quad n=2,3,\dots \, .
\label{dep}
\ee
In this case we define
\be
x = \half (d-2) \, \rho \, , 
\label{xrho}
\ee
and, with $v(\rho) = \hv(x)$, $\eta=0$,  \eqref{LPA3} becomes
\be
x \, \hv''(x)+ (\alpha+1-x)\, \hv'(x) + \frac{d}{d-2} \, \hv(x) = x \, \hv'(x)^2
\, , \qquad \alpha = \half N- 1 \, .
\label{LPA4}
\ee
Using \eqref{dep} it is easy to see that as $\vep\to0$ there is a solution
of the form
\be
\hv(x) = k_n \vep \, L^\alpha_n(x) + {\rm O}(\vep^2) \, ,
\label{solv}
\ee
for $L^\alpha_n$ a Laguerre polynomial, satisfying $xL^\alpha_n{}''(x)
+ (\alpha+1-x) L^\alpha_n{}'(x) + n L^\alpha_n(x)=0$ with the orthogonality
condition
\be
\int_0^\infty \!\!\! \rmd x \; e^{-x}x^\alpha \, L^\alpha_r(x)L^\alpha_s(x)
= \frac{1}{r!} \, \Gamma(\alpha+1+r) \, \delta_{rs} \, .
\label{ortho}
\ee

To determine $k_n$ we consider
the scalar product of both sides of \eqref{LPA4} with $L^\alpha_n$ to
${\rm O}(\vep^2)$ giving
\be
\half(n-1)^2 k_n \int_0^\infty \!\!\! \rmd x \; e^{-x}x^\alpha \, L^\alpha_n(x)^2
= k_n{\!}^2 \int_0^\infty \!\!\! \rmd x \; e^{-x}x^{\alpha+1} \, 
L^\alpha_n(x)\, L^\alpha_n{}'(x)^2 \, .
\ee
This requires
\be 
k_n = (-1)^n \frac{(n-1)^2}{n\, n!} \, 
\frac{(\alpha+1)_n}{G^{(\alpha,0)}_{nnn}} \, ,
\ee
where we define
\be
G^{(\alpha,l)}_{rst} = \frac{(-1)^{r+s+t}}{\Gamma(\alpha+l+1)} 
\int_0^\infty \!\!\! \rmd x \; e^{-x}x^{\alpha+l} \, L^{\alpha+l}_r(x) \, 
L^{\alpha+l}_s(x)\,  L^\alpha_t(x) \, .
\ee
It is straightforward to set up a perturbation expansion in $\vep$ for the
higher order terms in the solution \eqref{solv} as a series summing over 
Laguerre polynomials $L^\alpha_r$.

For the associated eigenfunctions and eigenvalues then letting
\be
\lambda_{l,m} = d - (d-2) \big (m +\half l + {\hat \lambda}_{l,m} \big ) \, ,
\ee
from  \eqref{eigf} and \eqref{Dop} we require that $ {\hat \lambda}_{l,m}$
is determined by
\be
{\hat\D}_l \hf_{l,m}(x) =  - \big ({\hat \lambda}_{l,m} + m \big ) \,
\hf_{l,m}(x) \, ,
\ee
where, with $\eta=0$ and $\alpha$ as in \eqref{LPA4}, $\D_l = (d-2){\hat \D}_l$,
\be 
{\hat \D}_l = x \frac{\rmd^2}{\rmd x^2} + (\alpha+l+1 -x)\frac{\rmd}{\rmd x}
- 2x \hv'(x) \frac{\rmd}{\rmd x} - l \, \hv'(x) \, .
\label{Dop2}
\ee
Note that we must have ${\hat \lambda}_{1,0}=0, \, {\hat \lambda}_{1,n-1}
= \vep(n-1)/(d-2)$ to ensure the exact results ${\lambda}_{1,0}= \half(d+2), \,
{\lambda}_{1,n-1}= \half(d-2)$.
As $\vep\to 0$ it is easy to see that
\be
\hf_{l,m}(x) \to L^{\alpha+l}_m(x) \, , \qquad 
{\hat \lambda}_{l,m} = {\rm O}(\vep) \, .
\ee
To first order, where we may use \eqref{solv} for $\hv$ in \eqref{Dop2}, 
perturbation theory gives
\begin{align}
{\hat \lambda}_{l,m} ={}&  k_n \vep \, \frac{m!}{\Gamma(\alpha+l+1+m)}
\int_0^\infty \!\!\! \rmd x \; e^{-x}x^{\alpha+l} \, L_m^{\alpha+l}(x)\Big (
2x L_n^\alpha{}'(x) \frac{\rmd}{\rmd x} + l \,  L_n^\alpha{}'(x) \Big) 
L_m^{\alpha+l}(x) \nn \\
={}& (-1)^n {n\, k_n} \, \frac{m!\, G^{(\alpha,l)}_{mmn}} {(\alpha+l+1)_m} \, \vep
= (n-1)^2 \, \frac{(\alpha+1)_n}{n!\, G^{(\alpha,0)}_{nnn}} \,
\frac{m!\, G^{(\alpha,l)}_{mmn}} {(\alpha+l+1)_m}\; \vep   \, .
\label{first1}
\end{align}

To lowest order $\eta={\rm O}(\vep^2)$. Using \eqref{e1} and \eqref{e2}
with \eqref{dep} and \eqref{solv} gives
\be
\eta \int_0^\infty  \!\!\! \rmd x \; e^{-x}x^{\alpha +1} =  
\frac{2n}{2n-1} \, \frac{k_n{\!}^2 \, \vep^2} {(n-1)^2} \bigg ( 
\int_0^\infty  \!\!\! \rmd x \; e^{-x}x^{\alpha+1} L^\alpha_n{}'(x)^2 + 
2\int_0^\infty  \!\!\! \rmd x \; e^{-x}x^{\alpha+2} L^\alpha_n{}''(x)^2 \bigg ) \, .
\ee
Noting that $ L^\alpha_n{}'(x)= -  L^{\alpha+1}_{n-1}(x)$ we then find from
\eqref{ortho}
\be
\eta = \frac{2}{n!} \bigg ( \frac{n k_n}{n-1} \bigg )^{\! 2} \, 
(\alpha+2)_{n-1} \, \vep^2 = 2(n-1)^2 n! \, 
\bigg ( \frac{(\alpha+1)_n}{n!^2\, G^{(\alpha,0)}_{nnn}}  \bigg )^{\! 2} \, \vep^2\, .
\label{eta1}
\ee

Explicit results in \eqref{first1} and \eqref{eta1} can be obtained using
\begin{align}
G^{(\alpha,l)}_{rst} = {}& \frac{(\alpha+l+1)_r \,
(\alpha+l+1)_s}{(l+r+s-t)!} \nn \\
& {}\times  \sum_n \frac{1}{n!(r-n)!(s-n)!} \,
\frac{(l+2n)!}{(t-r-s+2n)!} \, \frac{1}{(\alpha+l+1)_n} \, .
\label{LLL}
\end{align}
{}From \eqref{LLL} we have $G^{(\alpha,l)}_{00t} =  \binom{l}{t}$
so that $G^{(\alpha,1)}_{00t}=0$, $t\ge 2$, and 
$(\alpha+1)G^{(\alpha,1)}_{n-1\,n-1\,n} = \half n \, G^{(\alpha,0)}_{nnn}$. These
ensure that ${\hat \lambda}_{1,0}=0, \, {\hat \lambda}_{1,n-1} = \half (n-1)^2 \vep$ 
as required.
For the case of primary relevance here $n=2$ and we have
\begin{align}
k_2 = {}& \frac{1}{2(N+8)} \, , \qquad \eta = 2k_2{\!}^2(N+2)\, \vep^2 \, , \nn \\
{\hat \lambda}_{l,m} = {}& k_2 \big ( (l+2m)(l+2m-1) 
+ m ( N+2l+2m-2) \big ) \, \vep \, .
\end{align}
These results are in accord with those obtained \cite{Old,Phase} in early
calculations involving the $\vep$-expansion. For the potentially marginal
operators $\lambda_{4,0}= 2(N-4)k_2 \vep , \, \lambda_{2,1}=-14k_2 \vep$ and
$\lambda_{0,2} = - \vep$, so that the critical $N_c=4$ to leading order.

We may also extend the ${\rm O}(\vep)$ calculations to \eqref{Dh} for 
the non redundant derivative operators. With $\eta=0$ and the same change of
variables as in \eqref{xrho} this becomes
\begin{align}
\big ( {\hat \D}_1 - 2x \, \hv''(x) - 2 \hv'(x) \big ) \hh_m(x) = {}& 
- \big ( m + {\hat \lambda}_m \big ) \hh_m(x) \, , \nn \\
\lambda_m = - (d-2) \big ( m+1 +  {\hat \lambda}_m \big ) \, , & \qquad
m=0,1,\dots \, ,
\end{align}
where with \eqref{solv} we may take $\hh_m(x) = L_m^{\alpha+1}(x) + {\rm O}(\vep)$
and ${\hat \lambda}_m = {\rm O}(\vep)$. To first order perturbation
theory gives
\begin{align}
{\hat \lambda}_{m} = {}& {\hat \lambda}_{1,m} + 
2  k_n \vep \, \frac{m!}{\Gamma(\alpha+2+m)}
\int_0^\infty \!\!\! \rmd x \; e^{-x}x^{\alpha+1} \, L_m^{\alpha+1}(x)^2 \, \big (
x L_n^\alpha{}''(x)  +   L_n^\alpha{}'(x) \big) \nn \\
={}& {\hat \lambda}_{1,m} + (-1)^n {2 k_n \vep } \, \frac{m!} {(\alpha+2)_m} 
\big ( (n+\alpha) \, G^{(\alpha,1)}_{mm\,n-1} 
- \alpha \, G^{(\alpha+1,0)}_{mm\,n-1}\big ) \, ,
\label{first2}
\end{align}
using the identity $x L_n^\alpha{}''(x)  +   L_n^\alpha{}'(x)  = 
- (n+\alpha) \,  L_{n-1}^\alpha{}(x) + \alpha \, L_{n-1}^{\alpha+1}{}(x)$. 
When $n=2$ this gives
\be
{\hat \lambda}_{m} = {\hat \lambda}_{1,m} + k_2 \vep ( N+8m+2) \, , \quad
{\hat \lambda}_{1,m} = k_2 \vep \, m( N+ 6m+2) \, ,
\ee
giving
\be
{\hat \lambda}_{m} = \frac{(m+1)N+6m^2+10m+2}{2(N+8)}\, \vep \, .
\label{resRG}
\ee

\section{Perturbative Discussion}

The crucial significance of the $\vep$-expansion is that it is possible to
use standard perturbative methods in quantum field theory involving a Feynman
graph expansion. To show the parallel with the above treatment we describe
how the same results are found in terms of conventional $\beta$-functions.
For a general scalar lagrangian 
\be
\cL = \half \pr^\mu \phi \cdot \pr_\mu \phi + V(\phi) \, ,
\label{lag}
\ee
then we may define a  generalised $\beta$-function for the 
potential $V$, which is a linear function of the couplings\footnote{For 
$V(\phi)= \sum_I g^I \O_I$, where $\O^I$ form a basis of monomials in $\phi$, 
then $\beta_V(\phi) = \sum_I  \beta^I(g)\O_I$. Subsequently $\beta_V \cdot 
\frac{\pr} {\pr V} = \sum_I \beta^I(g) \frac{\pr}{\pr g^I}$.}, where
\be
B_V(\phi) =  \half (d-2)\phi \cdot \pr V(\phi) - d V(\phi) + \beta_V(\phi) \, .
\label{bbh}
\ee
The perturbative RG flow equations  are then 
\be
{\dot V}(\phi) = - B_V(\phi) \, ,
\ee
and the fixed points $V=V_*$ are hence defined by
\be
B_{V_*}(\phi) = 0 \, .
\label{fixp}
\ee

With $d$ as in \eqref{dep} then for a renormalisable theory $V(\phi)$
is a polynomial of degree $2n$.
Using background field methods and minimal subtraction \cite{JO} then in a 
perturbative expansion calculations give $\beta_V$ in the form
\be
\beta_V(\phi) = (\gamma_\phi \phi)\cdot \pr V(\phi) + {\tilde \beta}_V(\phi) \, ,
\label{betaV}
\ee
where $\gamma_{\phi,ij}$ is the anomalous dimension matrix for the field 
$\phi$ and ${\tilde \beta}_V(\phi)$ depends on $\phi$ solely in terms of 
scalar contractions of various products of $V_{i_1 i_2 \dots i_k}(\phi)=
\pr_{i_1} \dots \pr_{i_k} V(\phi)$ with $k\ge 2$.  For renormalisable $V$,  
$\gamma_\phi$ depends only on the $\phi$-independent
$g_{i_1 i_2 \dots i_{2n}} = V_{i_1 i_2 \dots i_{2n}}$ and 
$\beta_V(\phi)$ is also just a polynomial of degree $2n$.
In general there are contributions to $\beta_V$ at  $(p-1)(n-1)$ loops,
$p=2,3\dots$, when ${\tilde \beta}_V = {\rm O}(\pr^{2n(p-1)}V^p)$ and, if
$p>2$, $\gamma_\phi(g) =  {\rm O}(g^{p-1})$.

Assuming $V_*(\phi) = \frac{1}{(2n)!}g_{*\,i_1 i_2 \dots i_{2n}}
\phi_{i_1} \cdots \phi_{i_{2n}}$ then \eqref{fixp} determines, for 
$n=2,3,\dots$, $g_{*\,i_1 i_2 \dots i_{2n}}$ perturbatively as an expansion 
in $\vep$. At the fixed
point the anomalous dimension matrix for $\phi$ also defines
\be
\eta_{ij} = 2 \gamma_{\phi,ij} \big |_{g=g_*} \, .
\label{geta}
\ee

In the vicinity of a fixed point defining, for $F(\phi)$ an arbitrary 
polynomial of degree $2n$, a linear operator $\gamma$ by
\begin{align}
\beta_{V+F}(\phi) = {}& \beta_V(\phi) + \gamma F(\phi) + {\rm O}(F^2) \, , \
\label{defD}
\end {align}
then critical exponents are determined by the eigenvalue equation
\be
\Delta F(\phi) = 
\half  (d-2)\phi  \cdot \pr F(\phi) + \gamma_* F(\phi) = 
( d - \lambda) F(\phi) \, , \qquad \gamma_* = \gamma \big |_{V=V_*} \, .
\label{scaleF}
\ee
As in the discussion of the exact RG equations there are explicit 
eigenfunctions relating to $\phi$ itself. Corresponding to \eqref{betaV} we 
have
\be
\gamma = (\gamma_\phi \phi)\cdot \pr + {\tilde \gamma} \, ,
\ee
where ${\tilde \gamma}$ involves at least second order $\phi$-derivatives. 
Hence
\be
\Delta \phi_i  = \half(d-2) \phi_i + \half \eta_{ij} \phi_j \, ,
\label{phim}
\ee
and also by differentiating \eqref{fixp}
\be
\Delta V_{*,i}(\phi) =  \half(d+2) V_{*,i}(\phi) - \half \eta_{ij} V_{*,j}(\phi) \, .
\ee
$\cL$ is arbitrary up to total derivatives so in this discussion mixing with
operators which are just spatial derivatives is neglected. The relevant matrix
is triangular so the determination of scale dimensions is not affected.
Up to a derivative operator proportional to $\pr^2 \phi_i$, 
$V_{*,i}(\phi) \sim 0$, as a consequence of the field equations so this is 
redundant.

At lowest order, as shown using background field techniques with $n-1$ loop
Feynman diagrams in \cite{DO},
\be
\beta_V(\phi)^{(n-1)} = 
a_n V_{i_1 \cdots i_n}(\phi) V_{i_1 \cdots i_n}(\phi) \, ,
\label{beta1}
\ee
where higher loops are ${\rm O}(V^3)$. Also, for $2(n-1)$ loops,
\be
\gamma_{\phi,ij}{\!}^{(2n-2)} = 2\, \frac{(n!)^2}{(2n)!} \, a_n{\!}^2 \, 
g_{i\, i_1 i_2 \dots i_{2n-1}} \,g_{j \,i_1 i_2 \dots i_{2n-1}}  \, ,
\label{gam1}
\ee
where
\be
a_n = \frac{1}{(4\pi)^n}\, \frac{n-1}{n!} \, \Gamma \Big ( \frac{1}{n-1}
\Big )^{\! n-1} \, .
\label{an}
\ee
The definition \eqref{defD} and \eqref{beta1}, then determines 
$\D_V$ to lowest order
\be
{\tilde \gamma}^{(n-1)} = 2a_n\, V_{i_1 \cdots i_n}(\phi) F_{i_1 \cdots i_n} (\phi) \, .
\label{DV1}
\ee

As before we impose $O(N)$ symmetry so that
\be
V(\phi) = g \, \frac{1}{n!} (\half \phi^2)^n \, ,  \quad 
\beta_V(\phi) = \beta_g(g) \, \frac{1}{n!} (\half \phi^2)^n \, , \quad 
\gamma_{\phi, ij} = \gamma_\phi(g)  \, \delta_{ij} \, .
\label{Vstar}
\ee
At the fixed point \eqref{fixp}
\be
\beta_g(g_*) = (n-1)g_* \, \vep  \, , \qquad 
\eta = 2 \gamma_\phi(g_*) \, .
\label{fixg}
\ee
With $O(N)$ symmetry the eigenfunctions in \eqref{scaleF} have the form
\be
F_{l,m}(\phi) = ( t \cdot \phi )^l\, (\half \phi^2)^m \, , \quad t^2 =0 \, .
\label{Flm}
\ee
The corresponding eigenvalues are then
\be
d - \lambda_{l,m} = \half (d-2+\eta)(l+2m) + \gamma_{l,m} \, ,
\label{anom}
\ee
with $\gamma_{l,m}$ the anomalous dimension determined by ${\tilde \gamma}_* 
F_{l,m}(\phi) = \gamma_{l,m}  F_{l,m}(\phi)$.

To handle the combinatorics involved in evaluating \eqref{DV1} with $F$ as in 
\eqref{Flm} we re-express this using
\be
V_{i_1 \cdots i_n}(\phi) F_{i_1 \cdots i_n} (\phi) = 
\Big ( \frac{\pr}{\pr \phi} \cdot \frac{\pr}{\pr \phi'}
\Big )^{\! n} \big ( V(\phi) \, F(\phi') \big ) \Big |_{\phi'=\phi} \, ,
\ee
and then follow the method described in \cite{Phase}.  First we note
\be
(a\cdot \pr)^k (\half \phi^2)^n = \sum_r
\frac{k! \, n!}{r!\, (k-2r)! \, (n-k+r)!} \, (\half \phi^2)^{n-k+r}
(\half a^2)^r (a\cdot \phi)^{k-2r} \, ,
\ee
and then
\begin{align}
& \Big ( \frac{\pr}{\pr \phi} \cdot \frac{\pr}{\pr \phi'} \Big )^{\! k}
\big ( (\half \phi^2)^n \,  ( t \cdot \phi' )^l\, (\half \phi'{}^2)^m  \big )\nn \\
& \quad =  \sum_r \frac{k! \, n!}{r!\, (k-2r)! \, (n-k+r)!} \, (\half \phi^2)^{n-k+r} \,
(\phi \cdot \pr')^{k-2r} ( \half \pr'{}^2)^r \big ( ( t \cdot \phi' )^l 
(\half \phi'{}^2)^m  \big ) \, .
\end{align}
Using
\be
(\half \pr'{}^2)^r \big ( ( t \cdot \phi' )^l (\half \phi'{}^2)^m \big )
= \frac{m!}{(m-r)!} \, (\alpha+l+1+m-r)_r \, ( t \cdot \phi' )^l
(\half \phi'{}^2)^{m-r} \, , 
\ee
with $\alpha$ as in \eqref{LPA4}, and
\be
(\phi \cdot \pr')^p \big ( ( t \cdot \phi' )^l (\half \phi'{}^2)^m \big )
\big |_{\phi'=\phi} = 
\frac{(2m+l)!}{(2m+l-p)!} \, (t \cdot \phi )^l (\half \phi^2)^m \, ,
\ee
we then obtain\footnote{The coefficients $A_{kn,m}^{(\alpha,l)}$ satisfy
various identities, in particular
\begin{align}
(\alpha+l) & \big ( A_{k+1\,n,m}^{(\alpha,l)} + 2m(\alpha+l+m) A_{kn,m-1}^{(\alpha,l)}
\big )  \nn \\
= {}& m\,(2\alpha+l)(\alpha+l+m+n-k) \, A^{(\alpha,l+1)}_{kn,m-1} 
+ l\, (\alpha+l+m)(m+n-k) \, A^{(\alpha,l-1)}_{kn,m}  \, . \nn
\end{align}
When $N=1$, $A^{(-\frac{1}{2},l)}_{kn,m} 
= \frac{(2n)!\, (2m+l)!}{2^kn!(2n-k)!(2m+l-k)!}$ for $l=0,1$.}
\begin{align}
&\frac{1}{n!} \,
\Big ( \frac{\pr}{\pr \phi} \cdot \frac{\pr}{\pr \phi'} \Big )^{\! k}
\big ( (\half \phi^2)^n \,  ( t \cdot \phi' )^l\, (\half \phi'{}^2)^m  \big )
\Big |_{\phi'=\phi} = A^{(\alpha,l)}_{kn,m} \, 
( t \cdot \phi )^l  (\half \phi^2)^{m+n-k}   \, , \nn \\
& A^{(\alpha,l)}_{kn,m} =  \sum_r \frac{k! }{r!\, (k-2r)! \, (n-k+r)!} \,
\frac{m!}{(m-r)!} \, (\alpha+l+1+m-r)_r \, 
\frac{(2m+l-2r)!}{(2m+l-k)!} \, .
\label{comb}
\end{align}

Hence from \eqref{beta1} and \eqref{DV1} with \eqref{Vstar} and \eqref{Flm}
we may obtain to lowest order
\be
\beta_{g}(g)^{(n-1)} = A^{(\alpha,0)}_{nn,n} \; a_n \, g^2 \, ,
\label{bV1}
\ee
{}From the fixed point equation \eqref{fixg},  with $\eta =0$ and \eqref{bV1}, we get
\be
A^{(\alpha,0)}_{nn,n} \; a_n \, g_* = (n-1) \vep \, . 
\ee
In consequence
\be
{\tilde \gamma}_*^{(n-1)} F_{l,m}(\phi) = 
2a_n \, g_* \, A^{(\alpha,l)}_{nn,m} \; F_{l,m}(\phi) \, .
\ee
and, in \eqref{scaleF} and \eqref{anom}, to lowest order
\be
\gamma_{l,m} = 2 a_n \, g_* \, A^{(\alpha,l)}_{nn,m} 
= 2(n-1) \, \frac{A^{(\alpha,l)}_{nn,m}}{A^{(\alpha,0)}_{nn,n}} \, \vep \, .
\label{lambdap}
\ee
Note that $\gamma_{0,n} = \beta_g{\!}' (g_*)$.

As a special case of \eqref{comb} we also have
\be
\frac{\pr}{\pr \phi_i} \frac{\pr}{\pr \phi'{\!}_j}
\Big ( \frac{\pr}{\pr \phi} \cdot \frac{\pr}{\pr \phi'} \Big )^{\! 2n-1}
\big ( (\half \phi^2)^n \, (\half \phi'{}^2)^n  \big )
\Big |_{\phi'=\phi} = \half (2n)!n! \, (\alpha+2)_{n-1} \, \delta_{ij}
\ee
so that \eqref{geta} and \eqref{gam1} give
\be
\eta = 2 n! \, (\alpha+2)_{n-1} \, (a_n g_*)^2 
= 2(n-1)^2 n! \frac{1}{\big ( A^{(\alpha,0)}_{nn,n}\big)^2 }\; \vep^2   \, .
\label{etap}
\ee

The precise identity of the results \eqref{lambdap} and \eqref{etap} with those
obtained from the RG equation
\eqref{first1} and \eqref{eta1}, where ${\hat \lambda}_{l,m} = 
\half (n-1)\gamma_{l,m}$, follows from
\be
A^{(\alpha,l)}_{kn,m} = k!\, m! \, \frac{G^{(\alpha,l)}_{m+n-k\, mn}}
{(\alpha+l+1)_{m+n-k}} \, .
\ee

When $n=2$ the ${\rm O}(\vep)$ results may be read off from
\be
2a_2 g_* = \frac{\vep}{N+8} \, , \qquad
A^{(\alpha,l)}_{22,m} = (2m+l)(2m+l-1) + m ( N+2l+2m -2) \, .
\label{dfour}
\ee

\section{Mixing Effects}

For operators which are monomials $\phi^m$ with $m\ge 2n$ then perturbatively
it is necessary to include mixing effects with operators 
$(\pr\phi)^2\phi^{m-2n}$. For $m\ge 4n-2$ there is additional mixing with 
operators involving four derivatives, 
such as $(\pr^2\phi)^2$, but this is neglected here. We here
discuss how the treatment of the previous section may be extended and
show how reparameterisation invariance is manifest in a perturbative approach.

The initial renormalisable lagrangian $\cL$ in \eqref{lag} is extended to
\be 
\cL = \cL_V + \cL_{F,G} \, , \qquad \cL_V = 
\half \pr^\mu \phi \cdot \pr_\mu \phi + V(\phi) \, ,
\quad \cL_{F,G} =  
F(\phi) + \half G_{ij}(\phi) \pr^\mu \phi_i \pr_\mu \phi_j \, .
\label{LFG}
\ee
Although for general $F,G_{ij}$ the resulting $\cL$ is non renormalisable, 
keeping only counterterms which are linear in $F,G_{ij}$, we may consistently 
define a bare lagrangian $\cL_0$ which extends the renormalisable theory 
defined by $\cL_V$ to include first order perturbations by finite two 
derivative operators, so long as $F(\phi),G_{ij}(\phi)$ are constrained to 
avoid the necessity of four derivative counterterms.
As usual there are corresponding $\beta$-functions  
\begin{align}
B_F(\phi) = {}&  \half (d-2)\phi 
\cdot \pr F(\phi) - d F(\phi) + \beta_F(\phi) \, , \nn \\
B_{G,ij}(\phi) = {}& \half (d-2)\phi 
\cdot \pr G_{ij}(\phi) + \gamma_{\phi,ik}  G_{kj}(\phi)
+ \gamma_{\phi,jk}  G_{ik}(\phi)  + \beta_{G,ij}(\phi) \, , 
\label{BB}
\end{align}
which are linear in $F,G_{ij}$ so that
\be
\beta_F = \gamma_{FF} F + \gamma_{FG,ij} G_{ij} \, , \qquad
\beta_{G,ij} = \gamma_{GF,ij} F + \gamma_{GG,ijkl} G_{kl} \, .
\ee
Here $\gamma_{FF}, \gamma_{FG,ij}, \gamma_{GF,ij}, \gamma_{FG,ijkl}$ are 
differential operators depending on the renormalisable couplings or $V$,
clearly we have, restricted to $F(\phi)$ of degree $2n$, $\gamma_{FF} = \gamma$ 
as defined in \eqref{defD}. At a fixed
point the exponents are defined by the coupled equations
\be
B_F(\phi) \big |_{V=V_*} = - \lambda F(\phi) \, , \qquad B_{G,ij}(\phi) 
\big |_{V=V_*} = - \lambda G_{ij}(\phi) \, .
\label{exFG}
\ee

For the lagrangians in \eqref{LFG} we have
\be
\delta \cL_V = \delta_{F,G} \cL_{F,G} \quad \mbox{for} \quad
\delta \phi_i =  v_i(\phi) \, , \quad \delta \pr_\mu \phi_i = v_{i,j}(\phi)
\pr_\mu \phi_j \, ,
\ee
if\footnote{The full Lagrangian in \eqref{LFG} is invariant if \eqref{diffeo}
is extended to $\delta F = v \cdot \pr(V+F)$, $\delta G_{ij} = \pr_i v_j + \pr_j v_i
+ v\cdot \pr G_{ij}  + \pr_i v_{k}\, G_{kj} + \pr_j v_{k}\,G_{ik}$.}
\be
\delta_{F,G}  F(\phi) = v(\phi)\cdot \pr V(\phi)
\, , \quad  \delta_{F,G} G_{ij}(\phi) = \pr_i v_j(\phi) + \pr_j v_i(\phi) \, . 
\label{diffeo}
\ee
 If $F(\phi),G_{ij}(\phi)$ are restricted to ensure that no mixing with four
 derivative operators arises then it is necessary to require $v_i(\phi)
 ={\rm O}(\phi^{2(n-1)})$.

As was apparent in the discussion of renormalisation for general
two dimensional $\sigma$-models \cite{Hsigma,Back} invariance under
reparameterisations $\delta \phi_i=v_i(\phi)$ leads to a corresponding freedom
in the definition of the $\beta$-functions. Here we show how this leads to
relations for the exponents defined by \eqref{exFG}. Assuming first
\begin{subequations}
\begin{align}
\beta_F(\phi)\big |_{F = v\cdot \pr V, G_{ij} = \pr_i v_j + \pr_j v_i} = {}&
\big ( \gamma v(\phi) \big ) \cdot \pr V(\phi) 
+ v(\phi)\cdot \pr \beta_V(\phi) \, , \label{bdiffa} \\
\beta_{G,ij}(\phi)\big |_{F = v\cdot \pr V, G_{kl} = \pr_k v_l + \pr_l v_k} = 
{}& \pr_i \big ( \gamma v_j(\phi) - 2 \gamma_{\phi,jk} v_k(\phi) \big )
+ \pr_j \big ( \gamma v_i(\phi) - 2 \gamma_{\phi,ik} v_k(\phi) \big ) \, ,
\label{bdiffb}
\end{align}
\end{subequations}
with $\gamma$ defined in \eqref{scaleF}, then \eqref{BB} gives
\begin{subequations}
\begin{align}
B_F(\phi)  \big |_{F = v\cdot \pr V, G_{ij} = \pr_i v_j + \pr_j v_i} 
&  {} =  U(\phi) \cdot \pr V(\phi)  + v(\phi) \cdot \pr B_V(\phi) \, , 
\label{BBa} \\
B_{G,ij}(\phi) \big |_{F = v\cdot \pr V, G_{kl} = \pr_k v_l + \pr_l v_k} 
& {} =  \pr_i U_j(\phi) + \pr_j  U_i(\phi) \, ,
\label{BBb}
\end{align}
\end{subequations}
for
\be
U_i(\phi)=  \half (d-2)\phi \cdot \pr  v_i(\phi) + \gamma v_i(\phi)
- \half (d-2) v_i(\phi) - \gamma_{\phi,ij} v_j(\phi )  \, .
\label{defU}
\ee
A general justification of \eqref{BBa},\eqref{BBb} with \eqref{defU}
is described in appendix C.

At a critical point, where \eqref{fixp} holds, then for vector solutions of 
\eqref{scaleF},  $\Delta v_i = (d-\lambda_v) v_i$, there are, as a consequence
of \eqref{BBa},\eqref{BBb} corresponding solutions of  \eqref{exFG} such that
\be
\lambda = \lambda_v - \half (d+2 -\eta) \, , 
\label{red2}
\ee
assuming a diagonal form for $\eta_{ij}$. Thus in this perturbative context 
we reproduce \eqref{red}. In particular as a consequence of \eqref{phim}
we have the exact zero modes
\be
F_0 (\phi) = \phi \cdot \pr V(\phi) \, , \qquad
G_{0,ij} = 2 \delta_{ij} \, .
\ee

To verify these results we consider the lowest order perturbative results 
at $n-1$ loops
\begin{align}
\beta_F(\phi)^{(n-1)} ={}& 2 a_n V_{i_1 \dots i_n} (\phi)
\, F_{i_1 \dots i_n} (\phi) \nn \\
&{}- a_n \!\!\!\!\!
\sum_{\genfrac{}{}{0pt}{}{r,s,t\ge1}{r+s+t=n+1}} \!\!\!\! 
\frac{n!}{r!\,s!\,t!} \, {\hat K}_{rst} \,
V_{i_1 \dots i_r k_1 \dots k_t} (\phi) V_{j_1 \dots j_s k_1 \dots k_t} (\phi)\,
G_{i_1j_1,i_2\dots i_r j_2 \dots j_s} (\phi) \nn \\
&{}+ a_n \!\!\!\!\!
\sum_{\genfrac{}{}{0pt}{}{r\ge 2,s,t\ge1}{r+s+t=n+1}} \!\!\!\! 
\frac{n!}{r!\,s!\,t!} \, {\hat K}_{rst} \,
V_{i_1 \dots i_r k_1 \dots k_t} (\phi) V_{j_1 \dots j_s k_1 \dots k_t} (\phi)\,
G_{i_1i_2,i_3\dots i_r j_1 \dots j_s} (\phi) \, , 
\label{bnF}
\end{align}
where $a_n$ is as in \eqref{an} and 
\be
{\hat K}_{rst} = \Gamma(\tfrac{1}{n-1}) \, 
\frac{\Gamma(\tfrac{n-r}{n-1}) \, \Gamma(\tfrac{n-s}{n-1})\,\Gamma(\tfrac{n-t}{n-1})}
{\Gamma(\tfrac{r}{n-1}) \, \Gamma(\tfrac{s}{n-1})\,\Gamma(\tfrac{t}{n-1})} \, .
\label{Krst}
\ee
When $r=1,s+t=n$, ${\hat K}_{rst} = 1$. Also
\begin{align}
\beta_{G,ij}(\phi)^{(n-1)} =2a_n \big ( &
V_{i_1 \dots i_n} (\phi) \, G_{ij,i_1 \dots i_n} (\phi) 
+ 2 V_{i_1 \dots i_n(i} (\phi) \, G_{j)i_1,i_2 \dots i_n} (\phi) \nn \\ 
&{} - V_{i_1 \dots i_n(i} (\phi) \, G_{i_1i_2,j)i_3 \dots i_n} (\phi) \big ) \, .
\label{bnG}
\end{align}
We may directly verify that \eqref{bnF} and \eqref{bnG} satisfy 
\eqref{bdiffa},\eqref{bdiffb}
with ${\tilde \gamma}^{(n-1)}$ given by \eqref{DV1} and also $\gamma_\phi=0$. 
At the order given in \eqref{bnG} there are no contributions involving $F$. At the 
next non zero order
\begin{align}
\beta_{G,ij}(\phi)^{(2n-2)} = {}& - 8 a_n{\!}^2 \, \frac{(n!)^2}{(2n)!} \;
g_{i_1 \dots i_{2n-1}(i} \, F_{j)i_1 \dots i_{2n-1}} (\phi) \nn \\
&{} + (2n-1) 4 a_n{\!}^2 \,\frac{(n!)^2}{(2n)!} \; 
g_{i_1 \dots i_{2n-2}\,k(i}\, 
g_{i_1 \dots i_{2n-2}\,l \, j)} \, G_{kl}(\phi) 
+ {\rm O}(\pr G )  \, ,
\label{b2nG}
\end{align}
which is also compatible with \eqref{bdiffa},\eqref{bdiffb} using \eqref{gam1} for 
$\gamma_\phi$.

For definite results we assume $O(N)$ symmetry as in \eqref{Vstar}. To first
order in $\vep$ the equations \eqref{exFG} decouple and we may write
the eigenvalue equation
\be
\half (d-2) \, \phi \cdot \pr G_{ij}(\phi) + 2a_n g_* \, \D G_{ij}(\phi)
= - \lambda \, G_{ij}(\phi) \, ,
\ee
where from \eqref{bnG} $ \D G_{ij}$ is given by
\begin{align}
\D G_{ij}(\phi) = &\frac{1}{n!} \,
\Big ( \frac{\pr}{\pr \phi} \cdot \frac{\pr}{\pr \phi'} \Big )^{\! n}
\big ( (\half \phi^2)^n \, G_{ij}(\phi')  \big ) \Big |_{\phi'=\phi} 
\!\!\!  + \frac{2}{n!} \,
\Big ( \frac{\pr}{\pr \phi} \cdot \frac{\pr}{\pr \phi'} \Big )^{\! n-1}\!\!
\big ( \pr_k \pr_{(i} (\half \phi^2)^n \, G_{j)k}(\phi')\big ) 
\Big |_{\phi'=\phi} \nn \\
&{}- \frac{1}{n!} \,
\Big ( \frac{\pr}{\pr \phi} \cdot \frac{\pr}{\pr \phi'} \Big )^{\! n-2}
\big ( \pr_k \pr_l \pr_{(i} (\half \phi^2)^n \, \pr'{\!}_{j)} G_{kl}(\phi')\big )
\Big |_{\phi'=\phi} \, . 
\label{comb2}
\end{align}
Clearly for $G_{ij}(\phi)= \delta_{ij}$ then $\lambda=0$. In general we
have
\begin{align} 
\D\big ( \phi_i \phi_j (\half \phi^2)^m \big ) = {}&
\frac{1}{m+1} \,\pr_{(i} \; \frac{1}{n!} \,
\Big ( \frac{\pr}{\pr \phi} \cdot \frac{\pr}{\pr \phi'} \Big )^{\! n}
\big ( (\half \phi^2)^n \, \phi'{\!}_{j)} (\half \phi'{}^2)^{m+1} \big ) 
\Big |_{\phi'=\phi} \nn \\
&{} - \frac{1}{m+1} \, \D\big ( \delta_{ij}(\half \phi^2)^{m+1} \big ) \, ,
\end{align}
so that from \eqref{comb}
\begin{align}
G_{ij}(\phi) ={}&  \delta_{ij} (\half \phi^2)^{m+1} + (m+1) \, 
\phi_i \phi_j (\half \phi^2)^m \, , \quad m=0,1,\dots \, , \nn \\
\Rightarrow \quad
\lambda_m = {}& - (d-2)(m+1) - 2a_n g_* A^{(\alpha,1)}_{nn,m+1} =
- (d-2)(m+1) - \gamma_{1,m+1} \, ,
\label{Glam}
\end{align}
as expected from the general relation \eqref{red2} with \eqref{anom}
and $\eta=0$. In general 
\be
\frac{1}{m+1} \, \D\big ( \delta_{ij}(\half \phi^2)^{m+1} \big ) 
= \alpha_m \, \delta_{ij}(\half \phi^2)^{m+1} + \beta_m \, 
\phi_i \phi_j (\half \phi^2)^m \, ,
\ee
and we then find another eigenvalue, in addition to \eqref{Glam}, which 
can ne expressed in the form 
\be
\lambda_m = - (d-2)(m+1) - \gamma_m \, , \qquad
\gamma_m = 2a_n g_* \big ( (m+1) \alpha_m - \beta_m \big ) \, .
\label{lnew}
\ee
To obtain more explicit results for this we apply \eqref{comb2} to obtain
\begin{align}
\D\big ( \delta_{ij}(\half \phi^2)^{m} \big ) = {} & A^{(\alpha,0)}_{nn,m}
\, \delta_{ij} (\half \phi^2)^{m} \nn \\
&{}+ 2 \big ( (m+1)A^{(\alpha,0)}_{n-1\,n,m} - 2mA^{(\alpha,1)}_{n-1\,n,m-1} 
-(\alpha+n) m A^{(\alpha,1)}_{n-2\,n-1,m-1} \big )\nn \\
\noalign{\vskip -2pt}
& \hskip 1cm{}\times \big ( \delta_{ij} (\half \phi^2)^{m} + m \, 
\phi_i \phi_j (\half \phi^2)^{m-1} \big) \nn \\
&{}+ \frac{2}{n!} \,
\Big ( \frac{\pr}{\pr \phi} \cdot \frac{\pr}{\pr \phi'} \Big )^{\! n-1}\!\!
\big ( (\half \phi^2)^n \, \pr'{\!}_i \pr'{\!}_j (\half \phi'{}^2)^m  
\big ) \Big |_{\phi'=\phi} \nn \\ 
&{}+ (\alpha+n)\, \frac{2}{(n-1)!} \,
\Big ( \frac{\pr}{\pr \phi} \cdot \frac{\pr}{\pr \phi'} \Big )^{\! n-2}\!\!
\big ( (\half \phi^2)^{n-1} \, \pr'{\!}_i \pr'{\!}_j (\half \phi'{}^2)^m  
\big ) \Big |_{\phi'=\phi} \, .
\end{align}
{}From this we may determine
\begin{align}
(m+1) \alpha_m - \beta_m = {}& A_{nn,m+1}^{(\alpha,0)} \nn \\
&{}+ \frac{4}{N} \, (m+1)(\alpha+m+1) \big ( A^{(\alpha,0)}_{n-1\,n,m} 
+(\alpha+n) A^{(\alpha,0)}_{n-2\,n-1,m} \big ) \nn \\
&{}- \frac{4}{N} \, m(\alpha+m+2) \big ( A^{(\alpha,2)}_{n-1\,n,m-1} 
+(\alpha+n) A^{(\alpha,2)}_{n-2\,n-1,m-1} \big ) \, .
\label{abm}
\end{align}
Combining \eqref{abm} with \eqref{lnew} gives the exponents for new non
redundant operators so long as $N\ge 2$, for $N=1$ the corresponding
eigenfunction vanishes. When $n=2$ we have
\be
(m+1) \alpha_m - \beta_m = (m+2) N + 6m^2 +12m+4 \, .
\ee
Hence combining this with \eqref{dfour} the anomalous dimensions of
these derivative operators at the fixed point, to first order in $\vep$,
are given by
\be
\gamma_m = \frac{(m+2) N + 6m^2 +12m+4}{N+8}\, \vep \, .
\label{pertZ}
\ee
This may be compared with $2{\hat \lambda}_m$ in \eqref{resRG} which
was obtained from the approximate derivative expansion for the Polchinski RG
equations. Although similar they are not identical. The perturbative results
in \eqref{pertZ} are of course the first terms in a well defined expansion
to any order in powers of $\vep$.

\section{Conclusion}

The status of the equations presented in this paper for extending the local
potential approximation to the Polchinski exact RG equation is unclear, in that
there is no consistent derivation and the resulting equations for $V,Z$ are
partially decoupled in that the $V$-equation lacks expected $Z$ contributions.
Nevertheless the LPA equation now involves $\eta$ which, in this respect, 
is similar to an 
approximation made in a treatment of exact RG equations in \cite{Tetradis}.
It remains to be seen whether the introduction of terms involving $Z$
into the Polchinski LPA equations is at all possible, while maintaining the
crucial property of reparameterisation invariance, and so allowing a well
defined determination of $\eta$. Some time ago Morris \cite{Morris3} obtained,
with a particular cut off function $K(p^2)$ proportional to a simple power of 
$p^2$, and for the exact RG equations applicable to the one particle 
irreducible functional $\Gamma$, a set of equations, in a derivative expansion, which are
invariant under global rescaling of the fields. To this extent 
reparameterisation invariance is preserved and there is consequently an 
exact zero mode so that $\eta$ may be determined unambiguously.
However these equations are highly nonlinear and hard to analyse. The associated
zero mode eigenfunction for these equations has not apparently been found 
explicitly in the literature. If a derivative
expansion is to be consistent then results should not change dramatically when
the LPA for $V$ is extended to first order in derivative operators to
a pair of  coupled equations for $V,Z$. However 
results for some eigenvalues obtained using the Morris equations differ
significantly \cite{Con}. It would be very desirable to understand more
analytically what features of the equations in \cite{Morris3} ensure 
reparameterisation invariance so that this crucial constraint might be
imposed more generally in derivative expansion RG equations. It might also 
be helpful, as was the case here, to
construct an appropriate scalar product for the eigenfunctions whose 
corresponding eigenvalues are the essential output of exact RG equations.

In general reparameterisation invariance, which is related to issues
of scheme independence, is akin to a gauge invariance of the exact
RG equations \cite{Latorre}. As is well known for gauge theories violating 
gauge invariance in some approximation can lead to unphysical conclusions.
Emphasising the consequences of reparameterisation invariance, and
the consequential presence of an exact zero mode in the RG flow equations,
might also be a useful criteria for optimisation in equations where
reparameterisation invariance is not automatic.

It is of course non trivial that the LPA captures the global aspects of RG 
flows between various possible fixed points in scalar field theories. The
calculated critical exponents are also of essentially the correct magnitude
since the results must agree with those from ordinary perturbation
theory for non derivative operators to first order in $\vep$ (strictly this
appears to have been shown only when maximal $O(N)$ symmetry is required, it
is presumably true for fixed points with lesser symmetry although a general
argument appears to be lacking). A natural
constraint for any derivative expansion is that this agreement should extend
to non redundant scalar operators with two derivatives. The discrepancy
between the anomalous dimensions for such operators given by \eqref{resRG}
and \eqref{pertZ} shows that the equations discussed here are not fully
satisfactory in that respect. Perhaps more general derivative expansion
equations can be obtained by using more input from perturbative results.

\bigskip
\noindent
{\bf Acknowledgements}
\medskip

HO would like to thank Claude Bervillier and Oliver Rosten for
valuable email correspondence. DET would like to thank Trinity 
College for a summer research studentship.
\vfil\eject

\appendix\section{Large $N$ Calculations}

Solutions of the LPA fixed point equations become more tractable in the
large $N$ limit \cite{Com2,Kub}. We show here how these results may be used
to obtain a corresponding value for $\eta$ using \eqref{deta} or \eqref{e1}
and \eqref{e2}. In the large $N$ limit we suppose $v,\rho = {\rm O}(N)$
and taking the derivative of \eqref{LPA3} then gives
\be
\big ( (d-2+\eta) \rho - N + 4 \rho v' \big ) v'' = (2-\eta) v' - 2v'^2 \, ,
\ee
or
\be
2v' (s-v') \frac{\rmd \rho}{\rmd v'} - (d-2s+4v') \rho = - N \, , \qquad
s = 1 - \half \eta \, .
\ee
Such first order linear equations are readily solved giving $\rho$ as a 
function of $v'$,
\be
\frac{(s-v')^{\frac{1}{2s}d + 1}}{v'{}^{\,\frac{1}{2s}d - 1}} \, \rho
= C - \half N
\int_0^{v'} \! \! \rmd u \; \bigg ( \frac{s-u}{u} \bigg )^{\frac{1}{2s}d}\, ,
\ee
with the integral extended by analytic continuation for $\frac{1}{2s}d>1$.
To ensure a smooth continuously differentiable  solution for all $\rho > 0$ 
we must set the constant of integration $C=0$ and then the large $N$ result
can be expressed as
\be
\rho = \frac{N}{d-2s} \, F \big ( 2,1;2-{\ts\frac{1}{2s}}d;
{\ts\frac{1}{s}} v ' \big ) \, ,
\label{hyper}
\ee
for $F$ a standard hypergeometric function.

In the large $N$ limit $\eta\to 0$ so we may set $s=1$. When $d=3$,
$F(2,1;\half;v')=0$ for $v'=-0.6349132$ and so $v(0)\sim 0.2116377 N$.
For general $d$ we may invert \eqref{hyper} giving
\be
v'(\rho) = \frac{(4-d)(d-2)}{4N} \, ( \rho - \rho_0) + 
{\rm O}((\rho-\rho_0)^2 ) \, , \qquad \rho_0 = \frac{N}{d-2} \, .
\label{asy}
\ee
In integrals obtained from the scalar product \eqref{prod3} we then
have
\be
\rho^{\frac{1}{2}N } e^{- \frac{1}{2}(d-2)\rho - 2v(\rho)} 
\approx C \, e^{- \frac{d-2}{2N} \, (\rho - \rho_0)^2} \, ,
\ee
for some constant $C$, and so the dominant contribution for large $N$
arises for $\rho \approx \rho_0$ where we may use \eqref{asy}. Hence
\be
\frac{\big \langle \hf_{1,\phi} , v'^2  \big \rangle_1}
{\big \langle \hf_{1,\phi} , 1  \big \rangle_1 }\approx
\frac{(4-d)^2(d-2)}{16 N} \, .
\ee
Furthermore in \eqref{e2} for $\rho \approx \rho_0$
\be
\rho \big ( \half (2-\eta) - v'(\rho) - 2\rho v''(\rho) \big ) \approx
\half N \, ,
\ee
so that \eqref{e2} gives
\be
\frac{\big \langle \hf_{1,\phi} , y  \big \rangle_1}
{\big \langle \hf_{1,\phi} , 1  \big \rangle_1 }\approx
\frac{d(d-2)(4-d)^2}{8(d+2)N} \, .
\ee
Hence from \eqref{e1} the leading large $N$ result for $\eta$ in
this analysis is determined to be
\be
\eta = \frac{3 d(d-2)(4-d)^2}{8(d+2)N} + {\rm O}(N^{-2}) \, .
\label{llN}
\ee
The exact leading order large $N$ result $\eta \sim
2 (4-d) \Gamma(d-1) \sin \pi \half (d-2) / ( \pi d \Gamma(\half d)^2N)$
is numerically close to \eqref{llN}, coinciding as $d\to 4$.

\section{Perturbative Calculations}

We here outline how the perturbative results \eqref{bnF} and \eqref{bnG}
are obtained following the background field methods, with a background
field $\vphi$ and dimensional regularisation, used in \cite{DO}.  
To obtain \eqref{bnG} we consider vacuum graphs which are ${\rm O}(VG)$
\begin{align}
W_1 =  \sum_{r\ge 2}& \frac{1}{2 r!} \int \rmd^d x_1 \, \rmd^d x_2 \,
\Big \{ G_{ij,i_1 \dots i_r}(\vphi_1) \pr^\mu \vphi_{1i}\pr_\mu \vphi_{1j} 
\, G_0(x_{12})^r \nn \\
\noalign{\vskip -4pt}
&{}+ 2r\, G_{ii_1,i_2\dots i_r}(\vphi_1) \,  \pr^\mu\vphi_{1i}  \pr_\mu
G_0(x_{12}) \, G_0(x_{12})^{r-1} \nn \\
&{}+ r(r-1)\, G_{i_1i_2,i_3 \dots i_r} (\vphi_1)\,  \pr^\mu G_0(x_{12}) 
\pr_\mu G_0(x_{12}) \,  G_0(x_{12})^{r-2}\Big \}  
V_{i_1 \dots i_r} (\vphi_2)   \, ,
\label{WZ1}
\end{align}
for $\vphi_{1i} = \vphi_i(x_1)$ etc., and $G_0(x)$ the basic $d$-dimensional
propagator $-\pr^2 G_0(x) = \delta^d(x)$. Using, with $d$ as in \eqref{dep} 
and $a_n$ as in \eqref{an},
\begin{align}
& G_0(x)^n \sim  \frac{2}{\vep} \, \frac{n! \, a_n}{n-1} \, \d^d(x) \, , \quad
\pr_\mu G_0(x) \,G_0(x)^{n-1} \sim {} 
\frac{2}{\vep} \, \frac{(n-1)! \, a_n}{n-1} \, \pr_\mu \d^d(x) \, , \nn \\
& \pr^\mu G_0(x)\pr_\mu G_0(x)   \, G_0(x)^{n-2} \sim
\frac{2}{\vep} \, \frac{(n-2)! \, a_n}{n-1} \; \pr^2 \d^d(x) \, .
\end{align}
Consequently the necessary counterterm to cancel the $\vep$-pole in 
\eqref{WZ1} is given by
\begin{align}
\cL_{1{\rm c.t.}}(\vphi) = \frac{1}{\vep} \, \frac{a_n}{n-1} \,
\pr^\mu \vphi_i \pr_\mu \vphi_j  \big ( &
V_{i_1 \dots i_n} (\vphi) \, G_{ij,i_1 \dots i_n} (\vphi)
+ 2 V_{i_1 \dots i_n\,i} (\vphi) \, G_{ji_1,i_2 \dots i_n} (\vphi) \nn \\
&{} - V_{i_1 \dots i_n\,i} (\vphi) \, G_{i_1i_2,ji_3 \dots i_n} (\vphi) \big ) \, ,
\label{ctG}
\end{align}
which directly gives \eqref{bnG}.

Correspondingly to obtain \eqref{bnF} we consider vacuum graphs which
are  ${\rm O}(V^2G)$
\begin{align}
W_2 = {}& -\frac{1}{2} \sum_{\genfrac{}{}{0pt}{}{r,s\ge 0}{t\ge 1}} 
\frac{1}{r!s!t!}
\int \! \rmd^d x_1 \, \rmd^d x_2 \, \rmd^d x_3 \,
G_{ij,i_1 \dots i_r \, j_1\dots j_s}(\vphi_1)
V_{ii_1 \dots i_r\, k_1 \dots k_t}(\vphi_2)
V_{jj_1\dots j_s\, k_1 \dots k_t}(\vphi_3) \nn\\
\noalign{\vskip -14pt}
& \hskip 3cm {}\times  \pr^\mu  G_0(x_{12}) \, G_0(x_{12})^r \,
\pr_\mu  G_0(x_{13}) \, G_0(x_{13})^s \, G_0(x_{23})^t  \nn \\
{}& -\frac{1}{2} \sum_{\genfrac{}{}{0pt}{}{r\ge 0}{s,t\ge 1}} 
\frac{1}{r!s!t!}
\int \! \rmd^d x_1 \, \rmd^d x_2 \, \rmd^d x_3 \,
G_{ij,i_1 \dots i_r \, j_1\dots j_s}(\vphi_1)
V_{ij i_1 \dots i_r\, k_1 \dots k_t}(\vphi_2)
V_{j_1\dots j_s\, k_1 \dots k_t}(\vphi_3) \nn\\
\noalign{\vskip -14pt}
& \hskip 3cm {}\times  \pr^\mu  G_0(x_{12})\pr_\mu G_0(x_{12}) \, 
G_0(x_{12})^r \, G_0(x_{13})^s \, G_0(x_{23})^t \, . 
\label{W2}
\end{align}
By considering the pole in 
$(x_{12}{\!}^2)^{-\lambda_3} (x_{13}{\!}^2)^{-\lambda_2} (x_{23}{\!}^2)^{-\lambda_1}$
when $\lambda_1+\lambda_2+\lambda_3=d$, assuming $\lambda_1,\lambda_2,\lambda_3<\half d$
so that there are no sub-divergencies, we find, for ${\hat K}_{rst}$ as in \eqref{Krst},
\begin{align}
\pr^\mu  G_0(x_{12}) \, G_0(x_{12})^r \,
\pr_\mu  G_0(x_{13})& \, G_0(x_{13})^s \, G_0(x_{23})^t  \Big |_{r+s+t=n-1}\nn \\
\sim {}&
\frac{2}{\vep} \, \frac{n!\, a_n}{n-1}\, \frac{{\hat K}_{r+1\, s+1\,t}}{(r+1)(s+1)} \,
\delta^d(x_{12})\delta^d(x_{13}) \, , \nn \\
\pr^\mu  G_0(x_{12}) \pr_\mu G_0(x_{12}) \, 
G_0(x_{12})^r & \, G_0(x_{13})^s \, G_0(x_{23})^t \Big |_{r+s+t=n-1} \nn  \\ \sim {}&
- \frac{2}{\vep} \, \frac{n!\, a_n}{n-1}\, \frac{{\hat K}_{r+2\, st}}{(r+2)(r+1)} \,
\delta^d(x_{12})\delta^d(x_{13}) \, .
\end{align}
Hence \eqref{W2} requires the counterterm
\begin{align}
\cL_{2{\rm c.t.}}(\vphi) = \frac{1}{\vep} \, \frac{a_n}{n-1} \,
& \bigg ( {} - \!\!\!\!
\sum_{\genfrac{}{}{0pt}{}{r,s,t\ge1}{r+s+t=n+1}} \!\!\!\! 
\frac{n!}{r!\,s!\,t!} \, {\hat K}_{rst} \,
V_{i_1 \dots i_r k_1 \dots k_t} (\vphi) V_{j_1 \dots j_s k_1 \dots k_t} 
(\vphi)\, G_{i_1j_1,i_2\dots i_r j_2 \dots j_s} (\vphi) \nn \\
&{} +   \!\!\!\!
\sum_{\genfrac{}{}{0pt}{}{r\ge 2,s,t\ge1}{r+s+t=n+1}} \!\!
\frac{n!}{r!\,s!\,t!} \, {\hat K}_{rst} \,
V_{i_1 \dots i_r k_1 \dots k_t} (\vphi) V_{j_1 \dots j_s k_1 \dots k_t} 
(\vphi)\, G_{i_1i_2,i_3\dots i_r j_1 \dots j_s} (\vphi) \bigg )  \, , 
\label{ctF}
\end{align}
which leads to \eqref{bnF}.

A consistency check may be obtained by considering \eqref{ctF} when
$G_{ij}$ is a constant. The result is then
$\cL_{2{\rm c.t.}}(\vphi) = - \frac{1}{\vep} \, \frac{na_n}{n-1}
\, V_{ii_1\dots i_{n-1}}(\vphi) V_{ji_1\dots i_{n-1}}(\vphi)G_{ij}$. 
This may also be obtained directly from lowest order counterterm for
$V$ using that the kinetic term in $\cL$ defines a metric $\delta_{ij}
+ G_{ij}$. A similar argument is sufficient to obtain the last line
of \eqref{b2nG}.

\section{Analysis with Dimensional Regularisation}

Using dimensional regularisation with minimal subtraction for a theory 
defined by \eqref{lag} and \eqref{LFG} the diffeomorphism invariance 
under \eqref{diffeo} extends to the regularised theory and we demonstrate
here the consequences for the perturbative $\beta$-functions. 
The bare lagrangian, including all counterterms involving poles in
$\vep$ necessary for finiteness to first order in $F,G_{ij}$, has the general form
\be
\cL_0 = \mu^{-\vep} \big ( \half \pr^\mu \phi\cdot Z \pr_\mu \phi + \V(\phi) + 
\F(\phi) + \GG_{ij}(\phi) \, \half \pr^\mu \phi_i \pr_\mu \phi_j \big ) \, ,
\ee
where $\mu$ is the regularisation mass scale and $\F, \GG_{ij}$ are
linear in $F,G_{kl}$ and $Z_{ij}=Z_{ji}$ depends only on 
$g_{i_1 \dots i_{2n}}$. The bare couplings and field $\phi_0$ are defined
so as to absorb all dependence on $\mu$. For the standard renormalisable
theory, when $\F, \GG_{ij}$ are zero and $\V(\phi)$ is just a polynomial
of degree $2n$, the $\beta$-functions and $\phi$-anomalous dimension 
are defined through
\begin{align}
& \Big ( - \vep - (\hga_\phi \phi) \cdot \frac{\pr}{\pr \phi}
+ \hbe_V \cdot \frac{\pr}{\pr V} \Big ) \V(\phi) = 0 \, , \nn \\
& \Big ( - \vep + \hbe_V \cdot \frac{\pr}{\pr V} \Big ) Z_{ij} = 
\gamma_{\phi,ik} Z_{kj} + \gamma_{\phi,jk} Z_{ki} \, , 
\end{align}
where in terms of the $\beta$-functions considered earlier
\be
 \hbe_V (\phi ) = - \half \vep \, \phi \cdot \pr V(\phi) + \vep V(\phi)
+ \beta_V(\phi) \, , \qquad \hga_{\phi,ij} = - \half \vep \, \delta_{ij}
+ \gamma_{\phi,ij} \, .
\label{hBB}
\ee
For the extended theory, keeping only contributions to first order in 
$F, G_{ij}$, we also have
\begin{subequations}
\begin{align}
& \Big ( - \vep - (\hga_\phi \phi) \cdot \frac{\pr}{\pr \phi}
+ \hD_\beta \Big ) \F(\phi) = 0 \, , \label{rga} \\
& \Big ( - \vep  - (\hga_\phi \phi) \cdot \frac{\pr}{\pr \phi}
+ \hD_\beta \Big ) \GG_{ij}(\phi) =
\gamma_{\phi,ik} \GG_{kj}(\phi) + \gamma_{\phi,jk} \GG_{ki}(\phi) \, ,
\label{rgb}
\end{align}
\end{subequations}
for
\be
\hD_\beta = \hbe_V \cdot \frac{\pr}{\pr V}  
+ \hbe_F \cdot \frac{\pr}{\pr F} 
+ \hbe_{G,kl} \cdot \frac{\pr}{\pr G_{kl}} \, .
\label{Dbeta}
\ee
Here $\hbe_F$ is related to $\beta_F$ just as $\hbe_V$ is to $\beta_V$
in \eqref{hBB} and $\hbe_{G,ij}(\phi) = -  \half \vep \, \phi \cdot \pr
G_{ij}(\phi) + \beta_{G,ij}(\phi)$. 

For transformations as in  \eqref{diffeo} invariance of the regularised
theory requires
\begin{subequations}
\begin{align}
\F(\phi)\big |_{F = v\cdot \pr V, G_{ij} = \pr_i v_j + \pr_j v_i} = {}&
\D_v \F(\phi) = \tv(\phi) \cdot \pr \V(\phi) \, , \label{barediffa} \\
\GG_{ij}(\phi)\big |_{F = v\cdot \pr V, G_{kl} = \pr_k v_l + \pr_l v_k} =
{}& \D_v \GG_{ij}(\phi) = \pr_i \tv_k(\phi) Z_{kj} 
+ \pr_j \tv_k(\phi) Z_{ki} \, ,
\label{barediffb}
\end{align}
\end{subequations}
for
\be
\D_v = (v \cdot \pr V) \cdot \frac{\pr}{\pr F} 
+ (\pr_k v_l + \pr_l v_k) \cdot \frac{\pr}{\pr G_{kl}} \, .
\label{Dv}
\ee
In \eqref{barediffa},\eqref{barediffb} $\tv_i(\phi) = v_i(\phi) + \dots $
where we may have higher order terms involving poles in $\vep$.

The crucial constraints arise from the consistency conditions between
\eqref{rga},\eqref{rgb} and \eqref{barediffa},\eqref{barediffb} giving
\begin{subequations}
\begin{align}
\big [ \D_v , \hD_\beta \big ]  \F(\phi) = 
{}& \tu(\phi) \cdot \pr \V(\phi) \, ,  \label{conDa} \\
\big [ \D_v , \hD_\beta \big ]  \GG_{ij}(\phi) = {}&
\pr_i \tu_k(\phi) Z_{kj} + \pr_j \tu_k(\phi) Z_{ki} \, , \label{conDb}
\end{align}
\end{subequations}
where
\be
\big [ \D_v , \hD_\beta \big ] = 
\big ( \D_v \hbe_F - v \cdot \pr \hbe_V \big )  \cdot \frac{\pr}{\pr F} 
+ \big ( \D_v \hbe_{G,kl}\big ) \cdot \frac{\pr}{\pr G_{kl}} \, , 
\label{DDvb}
\ee
and
\be
\tu_i(\phi) =  \Big ( (\hga_\phi \phi) \cdot \frac{\pr}{\pr \phi}
- \hD_\beta \Big ) \tv_i(\phi) - \hga_{\phi,ij} \tv_j(\phi) \, .
\label{Uone}
\ee

To relate $\tv(\phi)$ to $v(\phi)$ we require that all counterterms are
determined by $\cL_0$. To achieve this we assume
\be
\tv_i(\phi) = v_i \cdot \frac{\pr}{\pr F}\, \F(\phi) \, .
\ee
Using \eqref{Uone} this leads to
\be
\tu_i(\phi) =  \Big ( v_i \cdot  \frac{\pr}{\pr F}\, \hbe_F - \vep v_i
- \hga_{\phi,ij} v_j \Big ) \cdot   \frac{\pr}{\pr F} \, \F(\phi)
+  \Big ( v_i \cdot  \frac{\pr}{\pr F}\, \hbe_{G,kl} \Big ) \cdot
\frac{\pr}{\pr G_{kl}} \,  \F(\phi) \, .
\ee
Comparing \eqref{conDa},\eqref{conDb} and \eqref{DDvb} with
\eqref{barediffa},\eqref{barediffb} and \eqref{Dv} requires 
\begin{subequations}
\begin{align}
\D_v \hbe_F(\phi) - v(\phi) \cdot \pr \hbe_V(\phi)  = 
{}& u(\phi) \cdot \pr V(\phi) \, ,  \label{fina} \\
\D_v \hbe_{G,ij} (\phi) = {}&
\pr_i u_j(\phi) + \pr_j u_i(\phi)  \, , \label{finb}
\end{align}
\end{subequations}
for
\be
u_i(\phi) =  v_i \cdot  \frac{\pr}{\pr F}\, \hbe_F(\phi) - 
\vep v_i (\phi) - \hga_{\phi,ij} v_j(\phi) \, .
\label{finc}
\ee
These results are equivalent to \eqref{BBa},\eqref{BBb} with
\eqref{defU}.

\section{Singularities of Solutions of RG Equation}

The critical requirement for solving the RG equation \eqref{LPA2} is
that it is necessary to fine tune $k=v(0)$ to ensure there are no
singularities for any real positive $\rho$. Nevertheless there are
necessarily singularities elsewhere in the complex plane. As shown
in \cite{Con} it is of interest to determine the location of such 
singularities so as to allow the use of conformal mapping techniques. 
The structure of the differential equation determines that the
singularities are simple poles of the form
\be
v'(\rho) \sim \frac{1}{\rho_0 \, e^{i\alpha \pi} - \rho} \, ,
\label{sing}
\ee
where reality of the equation ensures that $\pm \alpha$ must both
give singularities unless $\alpha=1$. We restrict then $0<\alpha\le 1$.
The singularities are determined by numerically integrating along
lines of constant argument, for $v(0)=k$ and matching with \eqref{sing}.

When $d=3$ the only singularities that are found within the radius in which
the numerical solution is valid are on the negative real 
axis, $\alpha=1$. Choosing $\eta=0$ the values of $k$ for suitable $N$
are given in Table 3 and the position of the closest singularity to the origin
is given in Table 9.
\begin{table}[!h]
\begin{tabular}{|r|r|} \hline
$N$~&$\rho_0$~~\\
\hline
$1$&\hfil$2.862$\\
$2$&\hfil$2.836$\\
$3$&\hfil$2.871$\\
$4$&\hfil$2.954$\\
$10$&\hfil$3.871$\\
\hline
\end{tabular}\caption{Values of $\rho_0$ for $N=1,2,3,4,10$}
\end{table}
 
We have also considered  $d=\frac{5}{2}$ since there are then two solutions
of physical interest with the additional solution arising for
$d< 3$ and representing  the tricritical fixed point.
 
For the solution corresponding to the standard Wilson-Fisher fixed point
the singularities are again just on the negative real axis. The results
are in Table 10.
\begin{table}[!h]
\begin{tabular}{|r|r|r|} \hline
$N$~&$k$~~~~~~~&$\rho_0$~~\\
\hline
$1$&\hfil$0.252995579$&\hfil$1.349$\\
$2$&\hfil$0.800566594$&\hfil$1.116$\\
$3$&\hfil$1.587372474$&\hfil$1.014$\\
$4$&\hfil$2.451092846$&\hfil$1.002$\\
$10$&\hfil$7.725892940$&\hfil$1.275$\\
\hline
\end{tabular}\caption{Values of $\rho_0$ and $k$ for $N=1,2,3,4,10$}
\end{table}
For the tricritical case there are genuine complex singularities,
results are given in Table 11.
\begin{table}[!h]
\begin{tabular}{|r|r|r|r|} \hline
$N$~&$k$~~~~~~~&$\rho_0$~~~~&$\alpha$~~~\\
\hline
$1$&\hfil$-0.043027023$&\hfil$6.3636$&\hfil$0.5150$\\
$2$&\hfil$-0.132093526$&\hfil$6.4736$&\hfil$0.4766$\\
$3$&\hfil$-0.279532538$&\hfil$6.5380$&\hfil$0.4383$\\
$4$&\hfil$-0.499803930$&\hfil$6.5544$&\hfil$0.3993$\\
\hline
\end{tabular}\caption{Tricritical values of $k$, $\rho_0$ and $\alpha$ for 
$N=1,2,3,4$}
\end{table}
An illustrative numerical solution for $v'(\rho)$ compared with the
pure pole term in \eqref{sing} is shown in Figure 3.
\begin{figure}[h!]
\centering
\includegraphics[trim= 20mm 80mm 20mm 70mm, clip, scale=0.5]{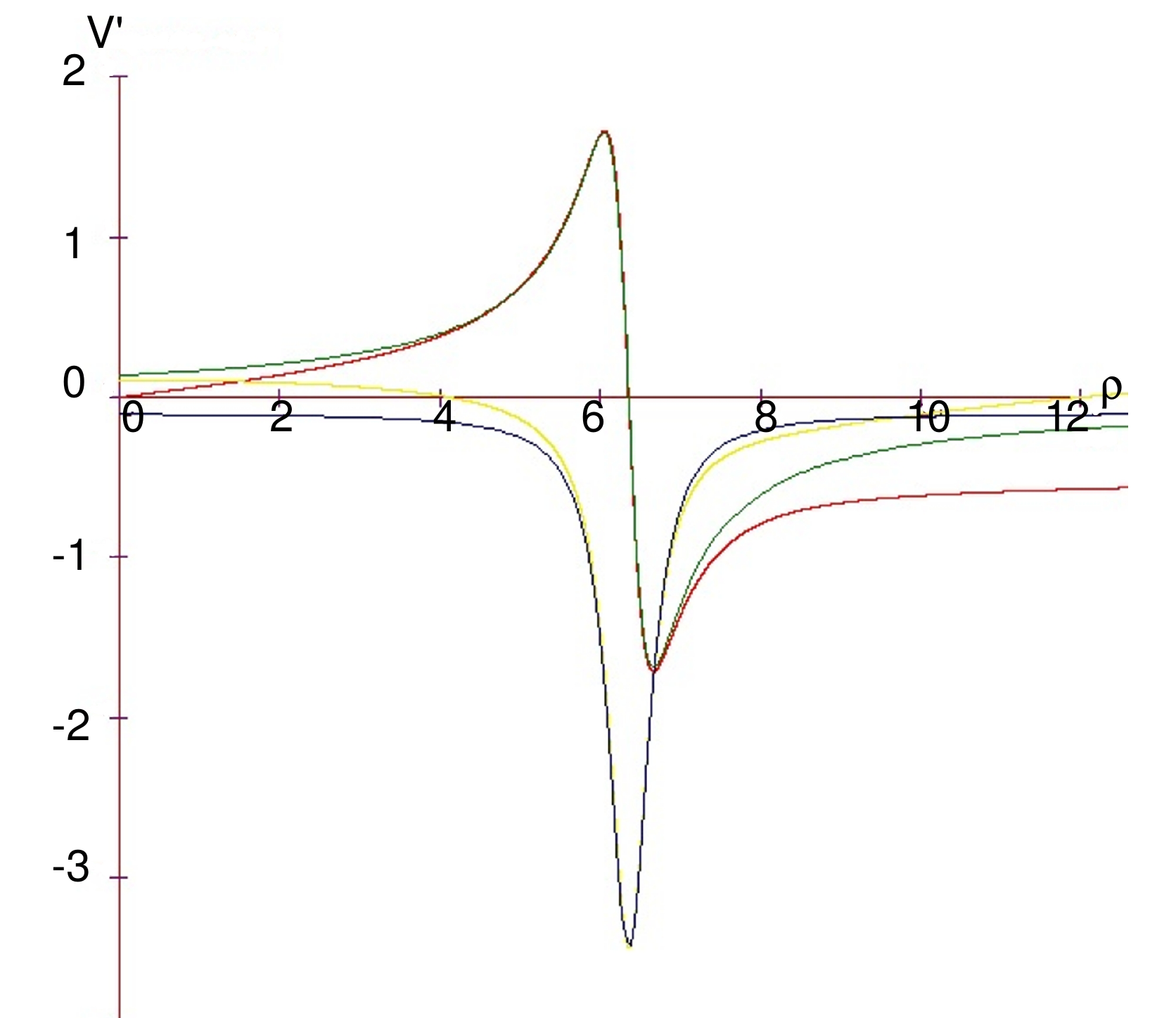}
\caption{$N=1$ tricritical solution for $v'(i \rho )$,
red and yellow lines are the real and imaginary parts, the green and blue
lines correspond to a pure pole as in \eqref{sing} with $\rho_0$ from Table 11}
\end{figure}

\renewcommand{\baselinestretch}{0.6}

\end{document}